\documentclass[12pt,preprint]{aastex}

\usepackage{natbib}
\bibpunct{(}{)}{;}{a}{}{,}

\slugcomment{Accepted in ApJ - March 30, 2007}

\shorttitle{The blast wave of Tycho's supernova remnant}

\shortauthors{Cassam-Chena{\"\i}, Hughes, Ballet and Decourchelle}

\begin{document}

\title{THE BLAST WAVE OF TYCHO'S SUPERNOVA REMNANT}

\author{Gamil Cassam-Chena{\"\i} and John P. Hughes}

\affil{Department of Physics \& Astronomy, Rutgers University \\
136 Frelinghuysen Road, Piscataway, NJ 08854, United States of
America}
\email{chenai@physics.rutgers.edu, jph@physics.rutgers.edu}

\and

\author{Jean Ballet and Anne Decourchelle}

\affil{UMR 7158, DSM/DAPNIA/SAp, CEA Saclay, 91191 Gif-sur-Yvette,
France} \email{jballet@cea.fr, adecourchelle@cea.fr}

\begin{abstract}
  We use the \textit{Chandra X-ray Observatory} to study the region in
  the Tycho supernova remnant between the blast wave and the shocked
  ejecta interface or contact discontinuity. This zone contains all
  the history of the shock-heated gas and cosmic-ray acceleration in
  the remnant. We present for the first time evidence for significant
  spatial variations of the X-ray synchrotron emission in the form of
  spectral steepening from a photon index of $\sim 2.6$ right at the
  blast wave to a value of $\sim 3.0$ several arcseconds behind. We
  interpret this result along with the profiles of radio and X-ray
  intensity using a self-similar hydrodynamical model including cosmic
  ray backreaction that accounts for the observed ratio of radii
  between the blast wave and contact discontinuity. Two different
  assumptions were made about the post-shock magnetic field evolution:
  one where the magnetic field (amplified at the shock) is simply
  carried by the plasma flow and remains relatively high in the
  post-shock region [synchrotron losses limited rim case], and another
  where the amplified magnetic field is rapidly damped behind the
  blast wave [magnetic damping case]. Both cases fairly well describe
  the X-ray data, however both fail to explain the observed radio
  profile. The projected synchrotron emission leaves little room for
  the presence of thermal emission from the shocked ambient
  medium. This can only be explained if the pre-shock ambient medium
  density in the vicinity of the Tycho supernova remnant is below $0.6
  \: \mathrm{cm}^{-3}$.
\end{abstract}


%*******************************************
\section{Introduction\label{sect-intro}}
%*******************************************

One of the most remarkable discoveries made by the \textit{Chandra
  X-ray Observatory} is that most of the X-ray synchrotron emission
from young ejecta-dominated supernova remnants (SNRs) is confined to
bright and geometrically thin rims at the blast wave behind which
there is little to no evidence for thermal X-ray emission from the
shocked ambient gas. Such thin X-ray synchrotron emitting rims are now
known to be common in young SNRs, having been detected in Cas A
\citep{hur00_CasA, gok01}, the Kepler SNR \citep{bay05}, the Tycho SNR
\citep{hwd02, wah05} and SN 1006 \citep{bay03, lor03, rob04}, among
others. The presence of X-ray synchrotron emission requires that the
SNR blast wave produces extremely high energy electrons ($> 1 \:
\mathrm{TeV}$), thereby providing strong support for efficient
particle acceleration at high Mach number shocks. The collisionless
shocks in supernova remnants are believed to be the main sites of
production and acceleration of Galactic cosmic rays (CRs), at least up
to the ``knee'' ($\sim 3000$ TeV) of the CR spectrum. However, the new
\textit{Chandra} findings pose two important questions that have not
yet been fully answered: (1) what is the underlying physical mechanism
for the featureless thin X-ray filaments, and (2) where is the thermal
X-ray emission from the shocked ambient medium?

Two interpretations have been proposed to explain the thin nonthermal
X-ray filaments seen in young SNRs: synchrotron cooling or magnetic
damping. In the first case, relativistic electrons accelerated at the
shock lose energy efficiently in the amplified post-shock magnetic
field and after advecting or diffusing a certain distance behind the
shock their synchrotron emission falls out of the X-ray band
\citep{ba03, vil03a, vob05, ba06, pam06}.  The other interpretation
posits that the filaments are regions where the magnetic field has
been amplified as well but where the rim widths are set by the damping
length of the magnetic field behind the shock \citep{poy05}.

Answering the first question posed above requires that we understand
the history or temporal evolution of both the distribution of the
accelerated particles and magnetic field at the shock and behind. In
fact, because the magnetic field strength determines the intensity of
the radiative losses and subsequent changes in the energy distribution
of the most energetic electrons, it is the most critical
ingredient. It is thus necessary to understand how the magnetic field
behaves at the shock (is it amplified?) and how it evolves downstream
(does it decrease and, if so, at what rate?). While theoretical
studies suggest that the magnetic field (specifically its turbulent
component) may be significantly amplified at the shock \citep{lub00,
  bel01, vle06}, it remains uncertain whether it is quickly damped on
a timescale shorter than that of the energy losses of electrons
\citep{poy05} or is simply carried along by the plasma flow in the
downstream region.  As we show below these two situations lead to
different predictions about the spectral variations of the X-ray
synchrotron emission and the relative radio and X-ray synchrotron
morphology at the blast wave. Our results will be based on a picture
in which the upstream magnetic field is a lot stronger than is
generally assumed for the interstellar medium. Assuming only
compression (by a factor of order 4), this allows us to simply model
the above mentioned strong magnetic field amplification.

The other important problem is the absence of thermal X-ray emission
from the shocked ambient medium in the region between the blast wave
and the contact discontinuity in young ejecta-dominated SNRs. Because
the blast wave also heats and compresses the ambient gas to X-ray
emitting temperatures, a thermal X-ray component is expected in
addition to the observed synchrotron emission. The strength of the
thermal component depends on both the density of the ambient medium
and the electronic temperature. It is unclear whether the lack of
X-ray emission from the shocked ambient medium reflects a density or a
temperature effect.

The absence of a complete theory on how collisionless shocks partition
their energy into bulk motion, thermal, and relativistic particles
does not allow us to calculate the post-shock electronic temperature
directly from dynamical quantities. If particle acceleration is very
efficient, the shock structure is modified with respect to the case
with no acceleration: the interaction region becomes much thinner
geometrically as well as cooler, which strongly influences the
resulting X-ray emission \citep{dee00, eld04, de05}. Further, even in
the absence of efficient shock acceleration, the electron temperature
can vary widely depending on whether the thermal electrons and ions at
the shock front share the same temperature value or the same velocity
distribution. In the latter case the electronic temperature will be
much less (by the mass ratio) than the ion temperature and the
different populations will gradually evolve toward equilibrium on the
timescale set by the particles exchanging energy through Coulomb
collisions. In fact the two cases just mentioned represent the extreme
ranges of a continuum of possible values that we parameterize in terms
of the electron-to-proton temperature ratio at the blast wave. There
is also a spatial variation in the X-ray intensity and spectrum of the
thermal emission from the shocked ambient medium. Right at the shock
front, the shock-heated gas has a low ionization state, which results
in little to no K-shell line emission in the X-ray spectrum. As the
gas evolves behind the shock, it becomes increasingly more ionized and
emission lines begin to appear. Setting a reliable constraint on the
ambient density based on the lack of the thermal X-ray emission
requires careful attention to the physical issues outlined above
(i.e., by understanding the history of the thermal particles in the
shocked ambient medium).

We have chosen to address the aforementioned questions using the
superb \textit{Chandra}\ X-ray data of the remnant of the supernova
event of 1572 that was observed by the Danish astronomer Tycho
Brah\'e. Now more than 430 years later, X-ray observations of the
Tycho SNR (hereafter, Tycho) show that the matter ejected from the
explosion (the ejecta) and heated by the reverse shock is distributed
over an almost circular shell with an angular radius of $\sim 240
\arcsec$. The recent advent of X-ray spectro-imagery has made it
possible to identify the major types of nucleosynthesis products and
study their spatial distribution.  This, in turn, constrains the
temperature distribution in the shocked ejecta, the level of mixing
and stratification between the different ejecta layers \citep{des01,
  hwd02}, the mechanism and energy of the explosion \citep{bab03,
  bab06} and allows us to quantify the development of Raleigh-Taylor
instabilities \citep[in the X-rays,][in the radio]{wah05, veg98}.

In fact, the recent high resolution X-ray observations not only show a
thick and clumpy shell of shocked ejecta but reveal the presence of
thin ($< 5 \arcsec$) and smooth rims preceding the ejecta by some ten
arcseconds (more or less), with very faint emission in between
\citep{hwd02, bay05, wah05}. These sharp rims of X-ray emission
demarcating the remnant's extent are naturally interpreted as tracing
the location of the blast wave. Similar rims are observed in other
historical SNRs such as Cas A, Kepler or SN 1006, but in Tycho, they
are much smoother and less structured which renders Tycho a perfect
target for the study of the rims. Their apparently featureless spectra
\citep{hwd02} and the failure of thermal models to reproduce the
observed radial profile of X-ray continuum emission suggest that most
of the X-ray radiation from the blast wave is due to synchrotron
emission by ultra-relativistic electrons \citep{cad04a, wah05, ba06}.

This is supported by radio observations with high angular resolution
($\sim 1\arcsec$) that show a morphology comparable to the X-ray image
as noted by \cite{acb94}, although the thin radio rims are not as
highly contrasted \citep{div91, rem97}. By comparison Tycho has a very
different appearance in the optical band. The optical filaments at the
rim show only H$\alpha$ emission from nonradiative shocks (there is no
radiative shock emission from either the ambient medium or ejecta) and
appear mostly in the east and north \citep{ghr00}. The study of the
most prominent optical knot on the eastern side yields a shock
velocity of $1900-2300 \: \mathrm{km/s}$, independent of distance
\citep{smk91, ghr01}. The electron-to-proton temperature ratio
associated with this knot was constrained to be less than $0.1$
\citep{ghr01}. The eastern and northern regions correspond to places
where the SNR is possibly interacting with cold and dense clouds as
suggested by H\textsc{i} and CO observations \citep{rev99,
  lek04a}. Radio \citep{rem97} and X-ray \citep{hu00} measurements
indicate an average expansion rate for Tycho of $0.11-0.12 \: \%
\mathrm{yr}^{-1}$, which corresponds to a shock velocity of $\sim 3300
\pm 1200 \mathrm{km/s}$ at an estimated, but still debated
\citep{scg95}, distance of $\sim 2.3 \pm 0.8 \; \mathrm{kpc}$
\citep{chk80, alt86, smk91, ru04}.

In Tycho, the observed closeness of the contact discontinuity to the
blast wave \citep{de05, wah05} and the observed concavity of the
synchrotron spectrum in the radio \citep{ree92} are suggestive of
efficient particle acceleration. Besides, Tycho being the remnant of a
thermonuclear explosion, its environment is expected to be uniform in
density and magnetic field. This uniformity is supported by the
quasi-circularity and regularity of the rims, and by the small
separation between the blast wave and the contact discontinuity, as
opposed to Cas A where the rims appear broken all around, and far from
the contact discontinuity, a result likely due to its expansion in a
stellar wind \citep{de05}. These characteristics all together make
Tycho a particularly well-suited remnant to investigate the problem of
the lack of thermal X-ray emission from the shocked ambient medium in
the context of efficient particle acceleration.

The goal of this paper is to address the above two important questions
on the origin of the filamentary nonthermal morphology of the blast
wave and lack of thermal X-ray emission from the shocked ambient
medium. For that purpose we investigate the radial variation of the
X-ray spectrum from the blast wave to the contact discontinuity (\S
\ref{subsect-PLindex} and \ref{subsect-sam}) and compare the rim
morphology in Tycho in the radio and X-ray bands (\S
\ref{subsect-radio-xray}). Results are interpreted and discussed with
the use of a cosmic-ray modified hydrodynamic model of SNR evolution
(\S \ref{sect-discussion}).

%*******************************************
\section{Data\label{sect-data}}
%*******************************************

\subsection{X-ray\label{subsect-data-xray}}

We used the \textit{Chandra} data of Tycho (\textsc{obs\_id} 3837)
observed on 2003 April 29 with the \mbox{ACIS-I} imaging spectrometer
in timed exposure and faint data modes. The X-ray analysis was done
using CIAO software (version 3.3). Standard data reduction methods
were applied for event filtering, flare rejection, gain
correction. The final exposure time amounts to 145 ks.

For the generation of images, spectra and PSF, we refer the reader to
the on-line guide of the \textit{Chandra} X-ray Center
website\footnote{See http://cxc.harvard.edu/ciao/threads/.} that we
followed for the current analysis. Spectra were always adaptively
grouped so that each bin has a signal-to-noise ratio greater than $5
\: \sigma$ and we used XSPEC\footnote{See
  http://heasarc.gsfc.nasa.gov/docs/xanadu/xspec/.} \citep[version
12.2.1,][]{doa03} for the spectral modelling.

\subsection{Radio\label{subsect-data-radio}}

We used the 1.4 GHz \textit{VLA} data of Tycho observed in 1994 and
1995 \citep{rem97}. These are the best currently available radio data
in terms of spatial resolution ($\sim 1\arcsec$).

Since the radio observations were done between 8 and 9 years before
the X-ray observations, we had to correct for the remnant's expansion
in order to compare both the radio and X-ray morphologies. For radial
profiles, we simply shifted the radio emission by a constant value of
$3\arcsec$. This corresponds approximately to an expansion rate of
$\sim 0.35\arcsec \: \mathrm{yr}^{-1}$ or $\sim 0.14\% \:
\mathrm{yr}^{-1}$ assuming a blast wave radius of $256\arcsec$. This
is roughly consistent with the average expansion rates of $0.113\% \:
\mathrm{yr}^{-1}$ in the radio \citep{rem97} and $0.124\% \:
\mathrm{yr}^{-1}$ in the X-ray \citep{hu00}. Note that a shift of the
radio profile by $4\arcsec$ would result in an even better match of
the sharp decline of the X-ray emission in various places around the
remnant but this corresponds to a larger expansion rate of $\sim
0.18\% \: \mathrm{yr}^{-1}$. Recently a second epoch \textit{Chandra}
observation of Tycho was approved in order to conduct a definitive
X-ray expansion study.

%*******************************************
\section{Results\label{sect-results}}
%*******************************************

\subsection{Radial variation of the synchrotron spectrum\label{subsect-PLindex}}

\subsubsection{Procedure}

\begin{figure*}[t]
\centering
\begin{tabular}{cc}
\includegraphics[width=8cm]{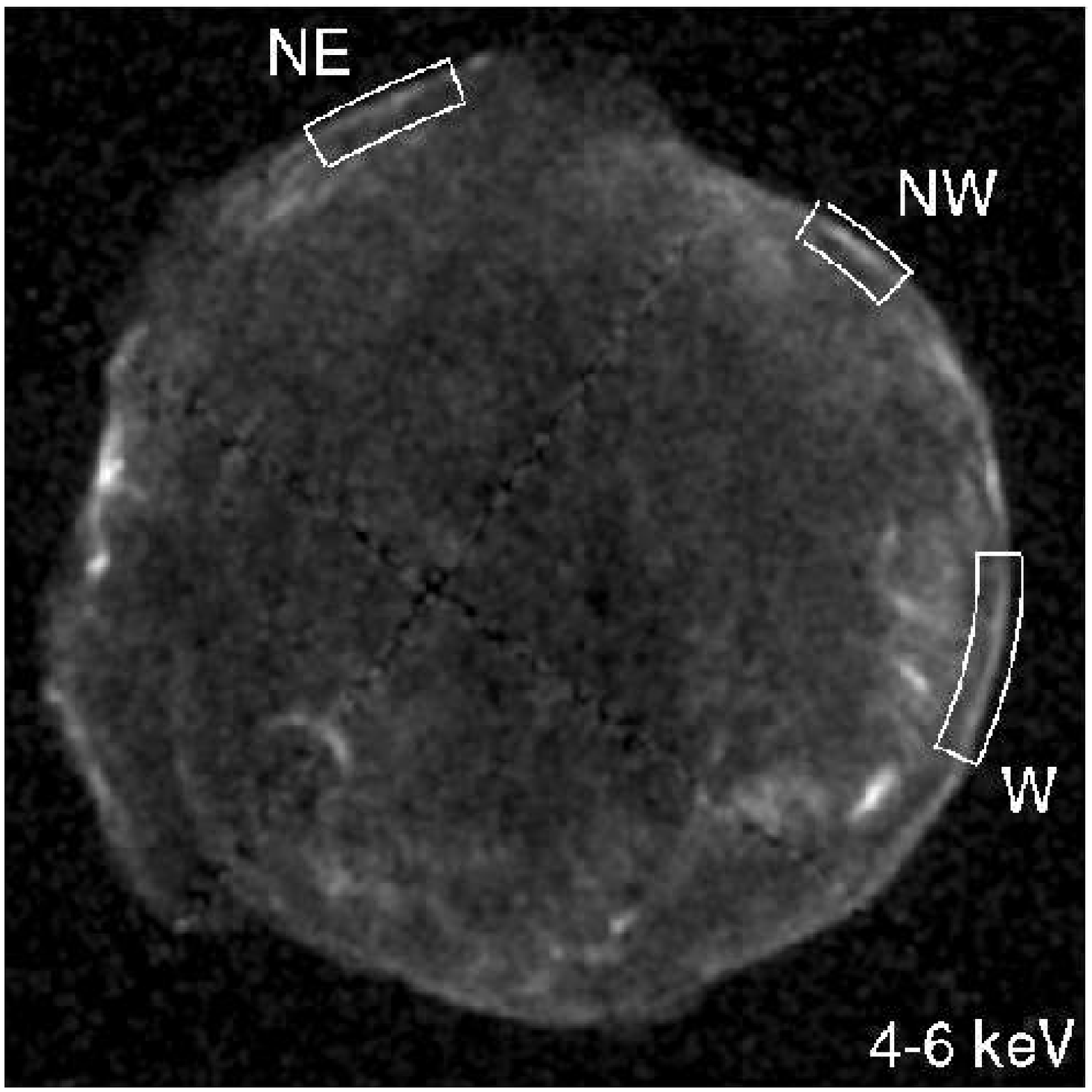} & \includegraphics[width=8cm]{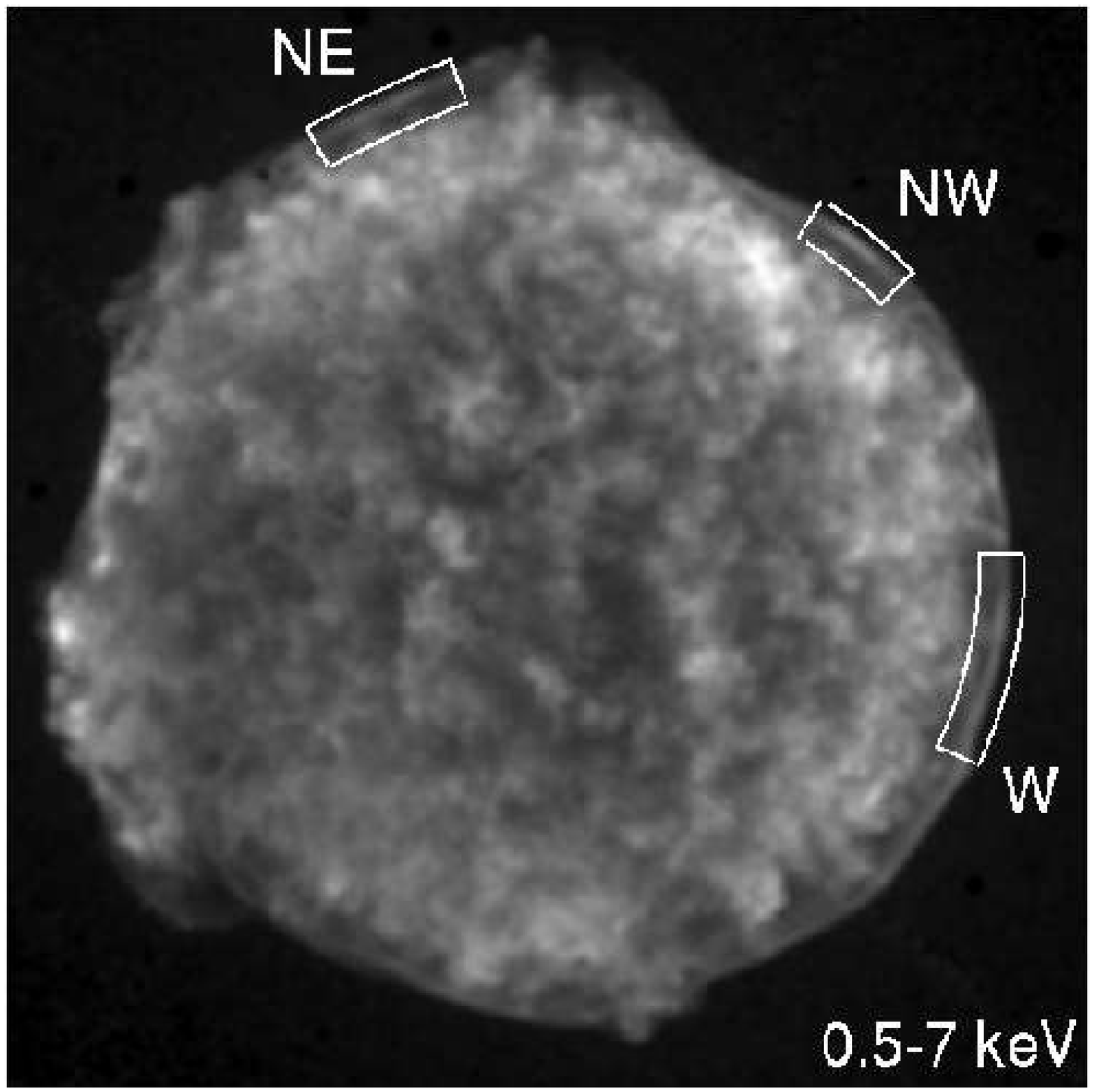} \\
\includegraphics[width=8cm]{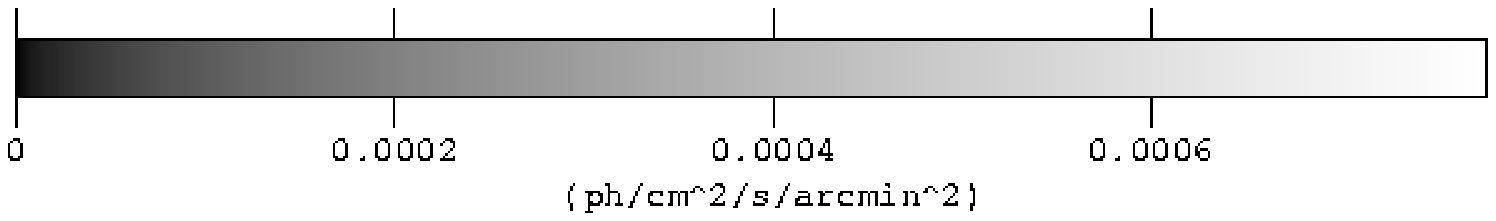} & \includegraphics[width=8cm]{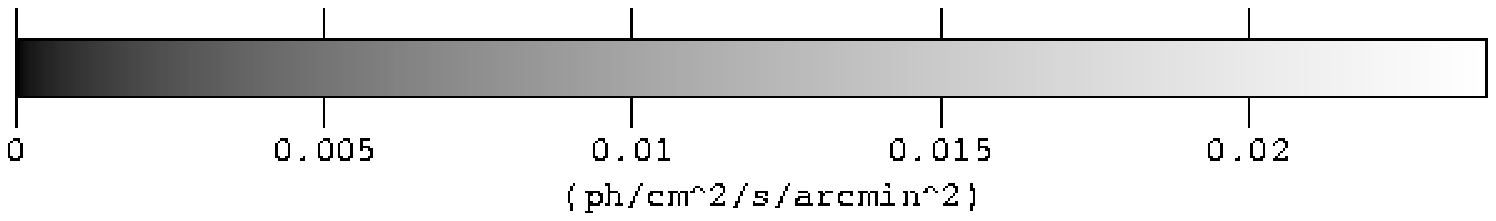}
\end{tabular}
\caption{\textit{Chandra} images of Tycho. \textit{Left panel}:
4-6 keV continuum band. \textit{Right panel}: 0.5-7 keV band. The
image of the continuum emission emphasizes the narrow rims
observed at the remnant's outer boundary. It is natural to
associate these rims with the blast wave. The broadband image
illustrates the closeness between the rims and the clumpy emission
from the shocked ejecta. The regions of interest are labelled as
W, NW, and NE (see Table \ref{tab-regions}). Both images are
corrected for exposure, vignetting and local astrophysical
background and are displayed with a square-root scaling.}
\label{fig_Tycho}
\end{figure*}

To search for radial spectral variations of the X-ray synchrotron
emission, we selected regions where the thin rims at the blast wave
are well defined and furthest from the shocked ejecta
emission. Regions with large gaps were identified using the results of
\citet{wah05}, while examining the 4-6 keV continuum band image (see
Fig.~\ref{fig_Tycho}) to avoid regions where overlapping filaments
occur at the rim. The regions are sectors chosen to have the largest
possible extent in azimuth to include as many photons as
possible. Radii were chosen to best follow the local curvature of the
rim. Figure \ref{fig_Tycho} shows the three regions that we selected
(W, NW and NE). Depending on the flatness of the rim, our approach
leads to vastly different sector radii of curvature. Table
\ref{tab-regions} (see label ``Rim'') shows for instance that the
resulting radius of curvature is far larger in the NE than in the W or
NW, because the rim is nearly straight in the NE. Note that these
local radii of curvature bear no relation to the radius of Tycho's
rim, which needs to be determined with respect to the remnant's global
center. The three azimuthally selected regions were divided radially
into thin sectors one arcsecond wide from which X-ray spectra were
extracted.

\begin{figure*}[t]
\centering
\includegraphics[bb=0 35 566 850,clip,width=13cm]{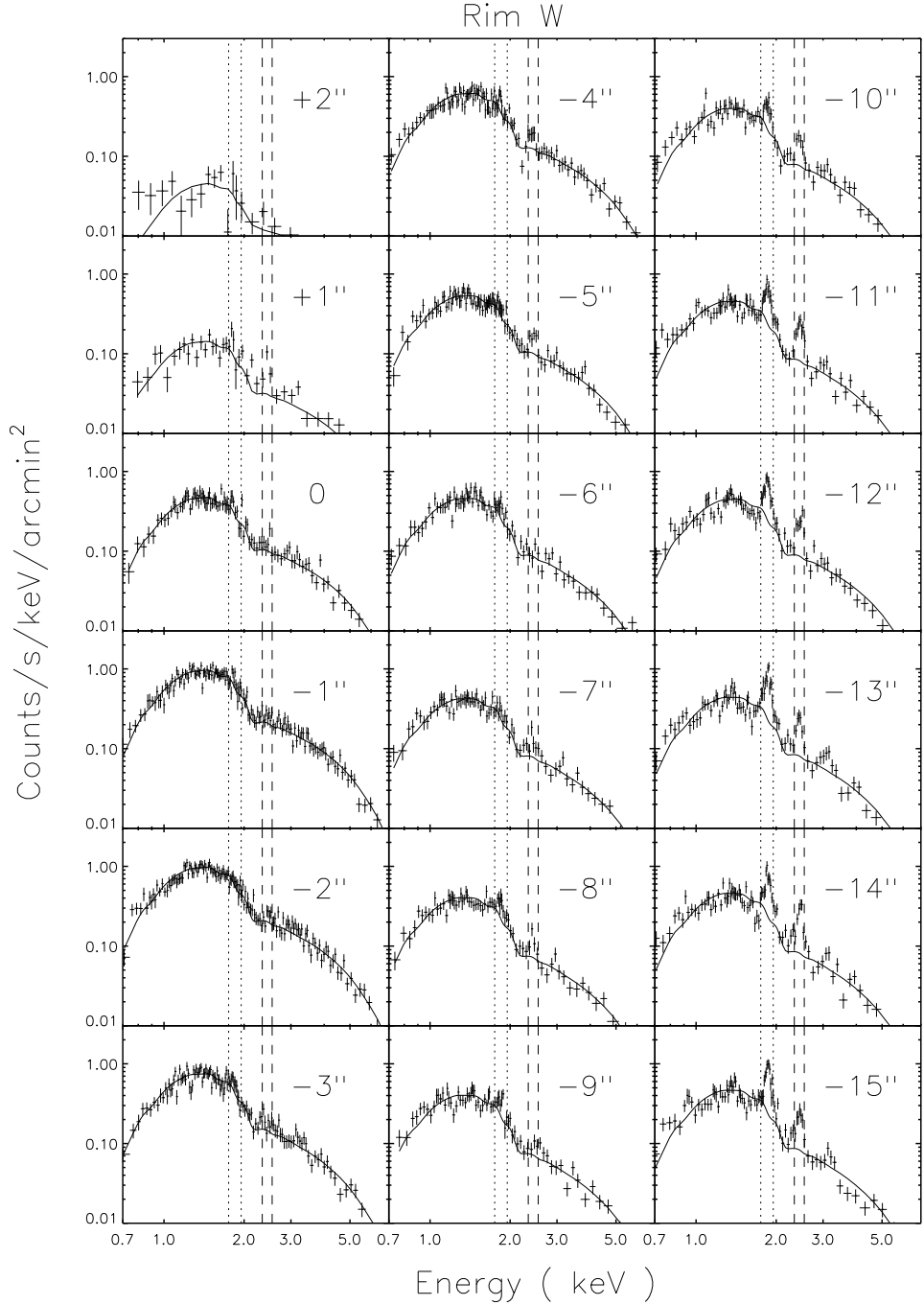}
\caption{Spectra extracted from the rim W (see Fig.
\ref{fig_Tycho}) fitted with a power-law model and a fixed
absorption of $0.7 \times 10^{22} \: \mathrm{cm}^{-2}$. The
numbers give the position from the blast wave in arcseconds. The
position 0 has been determined from an analysis detailed in \S
\ref{subsect-BW-CD} and its absolute location corresponds to a
sector whose inner and outer radii are $255\arcsec$ and
$256\arcsec$, respectively (see Table \ref{tab-regions}). The
dotted and dashed lines correspond to the Si \textsc{xiii}
He$\alpha$ ($1.75 - 1.94$ keV) and S \textsc{xv} He$\alpha$ ($2.34
- 2.55$ keV) bands, respectively.}
\label{fig_rim_spectra_grid_rimW}
\end{figure*}

The spectra obtained through this procedure are illustrated in Figure
\ref{fig_rim_spectra_grid_rimW}, which shows the evolution of the
X-ray spectrum for a particular region (rim W) as one moves in
radially from the outer boundary of the remnant toward the
interior. The first few X-ray spectra are characterized by a well
defined continuum which is sometimes associated with very faint
K-shell emission lines of silicon and sulfur. Then, as one moves
further in, these emission lines become more and more intense, until
they clearly stand out above the X-ray continuum. While the strong
emission lines can be naturally attributed to the shocked ejecta
\citep[see][]{des01, wah05}, it is not clear whether the faint lines
or other residual emission also come from the ejecta or whether they
arise from shocked ambient medium.

Since we are searching for spectral variations of the X-ray
synchrotron emission, we focus our attention on the
continuum-dominated regions. There are two possible origins for such
continuum: the thermal emission from the shocked ambient medium gas or
the nonthermal emission from the accelerated particles. Because the
morphology of the X-ray continuum emission and the quasi-featureless
nature of the overall rim spectra cannot be explained by a thermal
model in which most of the X-ray emission would come from the shocked
ambient medium \citep{cad04a, wah05, ba06}, we associate the continuum
with the synchrotron emission from relativistic electrons.

\subsubsection{Blast wave and contact discontinuity\label{subsect-BW-CD}}

Before proceeding to the spectral fits, we first describe how we
locate the fluid discontinuities. Determining the precise radial
position of the blast wave is somewhat delicate. Assuming spherical
symmetry, the blast wave radius is given by the position where the
emission from the remnant drops to zero (after background
subtraction). However, in practice one needs to account for the
instrumental point-spread-function (PSF), which causes the region
where the emission decreases to zero to broaden.

To take account of the PSF one needs to model the emission. We have
fitted a projected shell model with either uniform or exponential
emissivity profile convolved with the \textit{Chandra} PSF (extracted
at each particular rim), as done by \cite{wah05}, to X-ray synchrotron
brightness profiles (precisely, those that will be obtained in \S
\ref{subsect-pl+ej}). This allowed us to derive estimates of both the
blast wave position\footnote{Emission from a shock precursor could in
  principle alter the estimate of the blast wave radius but the X-ray
  synchrotron brightness profiles fitted with the convolved projected
  shell model allow little room for any precursor X-ray emission.} and
width of the shell in the W, NW and NE rims (see Table
\ref{tab-radii-psf}).  For uniform or exponential emissivity profiles,
we found an upper limit on the shell thickness of $\sim 0.6\arcsec$ in
the W rim with even lower values $< 0.4\arcsec$ in the NW and NE
rims. We note that the model profile for a thin spherical shell is
less peaked than the data in the NW and NE rims, suggesting some
deviation from our assumption of pure spherical symmetry. This
procedure resulted in an accurate determination of the blast wave
location for the three azimuthal regions.

We used the results of \cite{wah05} to determine the ratio of blast
wave to contact discontinuity radii and their uncertainties in the
three azimuthal regions.  The numerical values are $R_{s}/R_{c} \simeq
1.11$ in the W rim and $R_{s}/R_{c} \simeq 1.09$ in both rims NW and
NE with uncertainties of order of $3-5 \%$.  In the numerical results
we used $R_{s}/R_{c} \simeq 1.113$ $(=256\arcsec/246\arcsec)$ for the
W rim.  All these values are higher than the azimuthally averaged
value of $1.075$ quoted by \cite{wah05}. This is simply because we
initially chose portions of the X-ray rims that were farthest from the
contact discontinuity. In our analysis below we show how our results
change as a function of $R_{s}/R_{c}$.  Finally for completeness we
note that the contact discontinuity lies some $26\arcsec$, $20\arcsec$
and $18\arcsec$ behind the blast wave for the W, NW and NE rims,
respectively. In all cases the set of radial spectra we show do not
extend this far into the remnant's interior.

\subsubsection{Power-law model}

As a first approach to search for radial spectral variations of the
synchrotron emission, we used a phenomenological power-law to model
the different rim spectra ignoring line emission. While this is not
appropriate for the inner regions with strong emission lines (but
which may include a certain level of synchrotron emission due to
projection effects), it is a very good model for the quasi-featureless
outer regions (see Fig.  \ref{fig_rim_spectra_grid_rimW}).

\begin{figure*}[th]
\centering
\includegraphics[width=16cm]{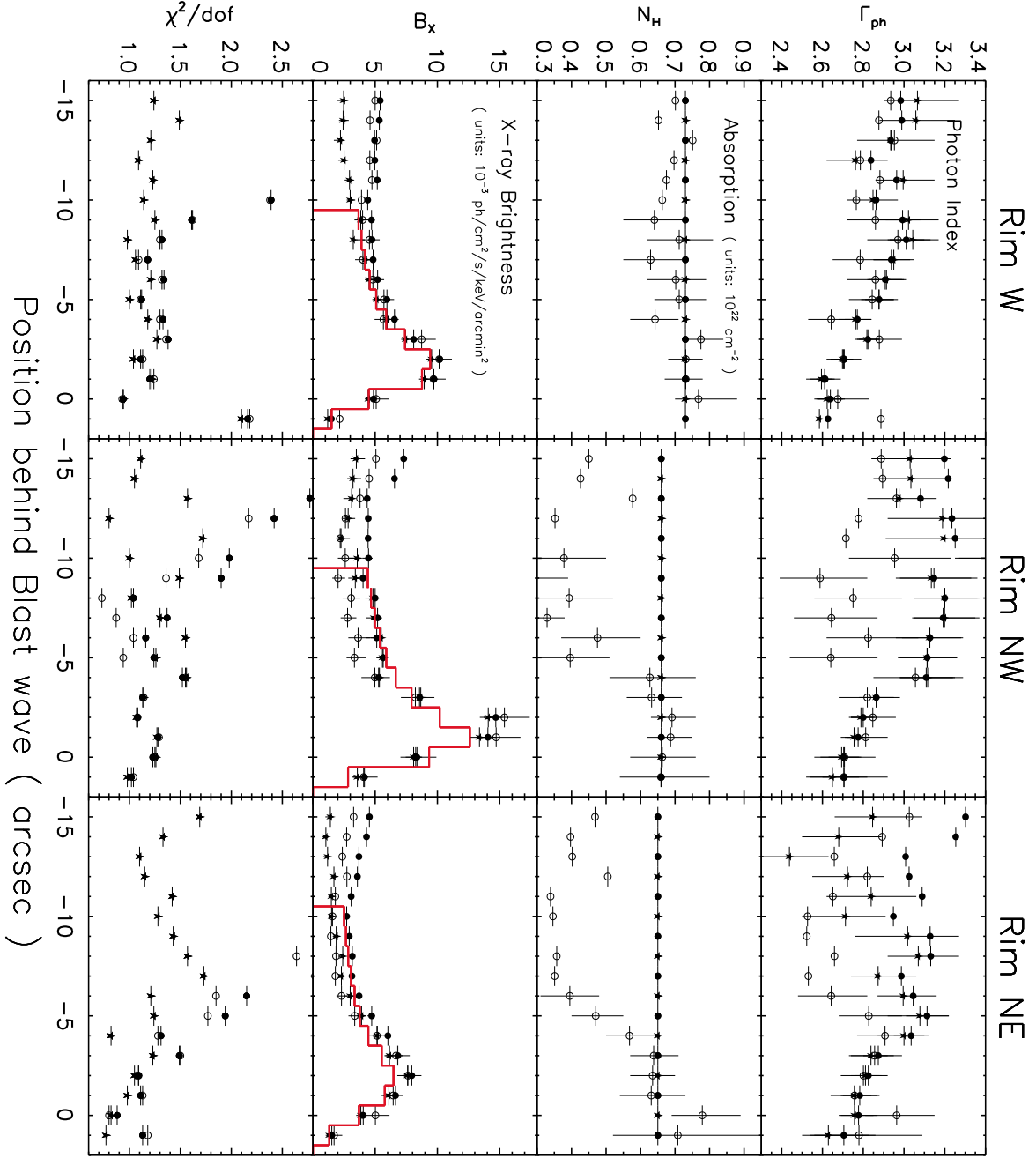}
\caption{Photon index $\Gamma_{\mathrm{ph}}$, X-ray absorption
$\mathrm{N}_{\mathrm{H}}$, X-ray brightness corrected for
absorption $\mathrm{B}_{\mathrm{X}}$ and reduced $\chi^2$ as a
function of position behind the blast wave in the rims W
(\textit{left panel}), NW (\textit{middle panel}) and NE
(\textit{right panel}) obtained for different spectral models:
power-law with free absorption ({\Large $\circ$}), power-law with
a fixed absorption ({\Large $\bullet$}), power-law plus a template
for the shocked ejecta with a fixed absorption ({\Large $\star$})
[see Table \ref{tab-slope-xspec-fit} for numerical values]. The
X-ray surface brightness profile obtained with this latter
spectral model was fitted until position $-9\arcsec$ with a
uniform-emissivity projected shell model convolved by the
\textit{Chandra} PSF (best-fit in red line). The errors are in the
range $\Delta \chi^2 < 2.7$ (90\% confidence level) on one
parameter and are given only when $\chi^2 / \mathrm{dof} < 2$. }
\label{fig_param_vs_radius_mo_all_one_fig}
\end{figure*}

Figure \ref{fig_param_vs_radius_mo_all_one_fig} shows the best-fit
parameters derived from a simple power-law model (data points with the
symbol {\Large $\circ$}) in the W, NW and NE rims. We show the
power-law index (top panel), the line-of-sight hydrogen column
density\footnote{ We used the solar abundance values from
  \cite{ane82} for the calculation of absorption (WABS model in
  XSPEC).} (second panel), the X-ray brightness (from the fitted
normalization of the power-law model at 1 keV) corrected for
interstellar absorption (third panel) and the best-fit reduced
$\chi^2$ (bottom panel) as a function of position behind the blast
wave. Starting from the blast wave, we see that the photon index
increases over a few arcseconds where the X-ray synchrotron brightness
is high. This is a first indication for radial variations of the
synchrotron spectrum in the rims. Then, as we move further in, the
photon index profile does not show a clear pattern anymore.

In fact, we note that the photon index and the absorption appear to be
highly correlated (see in particular the NW and NE rims).  Right at
the edge of the remnant where the emission is the brightest, the
absorption is highest. This is observed in three different
widely-spaced places around the remnant. Since there is no reason for
a sudden increase of the interstellar absorption precisely at the
position of the bright rims all around Tycho, we believe this spatial
variation to be spurious and related to an overly simplified spectral
model for the interior emission.  We therefore fixed the absorption to
a local average value for each azimuthal region determined by
averaging the best-fit absorption values between position 0 and
position $-4\arcsec$ where the rims are bright and featureless. This
yields $\mathrm{N}_{\mathrm{H}} = 0.73 \times 10^{22} \:
\mathrm{cm}^{-2}$ in the rim W, $0.66 \times 10^{22} \:
\mathrm{cm}^{-2}$ in the rim NW and $0.65 \times 10^{22} \:
\mathrm{cm}^{-2}$ in the rim NE.

Figure \ref{fig_param_vs_radius_mo_all_one_fig} shows the best-fit
parameters derived from a simple power-law model with a fixed
absorption (data points marked with {\Large $\bullet$}) in the W, NW
and NE rims. We observe an even more remarkable increase of the photon
index (top panel) over at least $\sim 5\arcsec$ behind the blast wave
where the regions have a quasi-featureless spectrum. In the three
rims, the photon index starts at $\sim 2.6-2.8$ at the shock and rises
up to $\sim 3.0-3.2$. The range in photon index variation is about
$0.3-0.4$. The photon index profiles then reach a more or less uniform
value further in.

\begin{figure}[t]
\centering
\includegraphics[width=8cm]{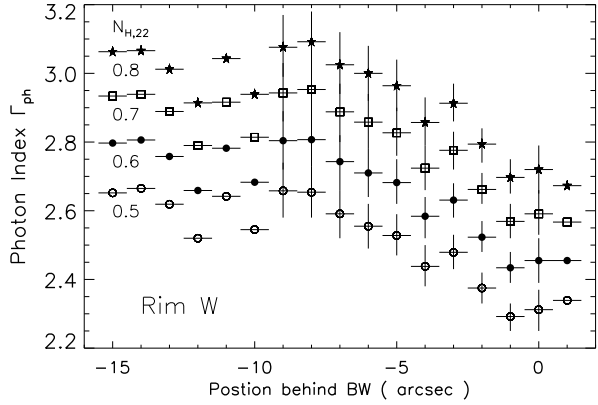}
\caption{Photon index $\Gamma_{\mathrm{ph}}$ obtained by fitting
the spectra of the rim W (see Fig.
\ref{fig_rim_spectra_grid_rimW}) with a power-law model and for
different values of a fixed absorption $\mathrm{N}_{\mathrm{H},22}
\equiv \mathrm{N}_{\mathrm{H}} / 10^{22} \: \mathrm{cm}^{-2}$ of:
$0.5$ ({\Large $\circ$}), $0.6$ ({\Large $\bullet$}), $0.7$
($\Box$) and $0.8$ ({\Large $\star$}). The errors are in the range
$\Delta \chi^2 < 2.7$ (90\% confidence level) on one parameter and
are given only when $\chi^2 / \mathrm{dof} < 2$. Varying the
absorption by $0.1 \times 10^{22} \: \mathrm{cm}^{-2}$ shifts the
photon index by $\sim 0.15$ but does not change the overall
profile. In the rim W, the best-fit is obtained for
$\mathrm{N}_{\mathrm{H},22}
 \simeq 0.7$. } \label{fig_slope_vs_radius_rimW_mowapo_nh}
\end{figure}

In Figure \ref{fig_slope_vs_radius_rimW_mowapo_nh}, we demonstrate
that the gradient of spectral index is robust to column density
variations. However, it is important to have a good estimate of the
interstellar absorption because it determines the absolute value of
the photon index, particularly important for broad-band non-thermal
models. The strongest constraint other than that from rim spectra
comes from the shocked ejecta X-ray emission (there is too much
uncertainty associated with converting H\textsc{i} data or optical
extinction values to X-ray column densities).  Therefore, we have
fitted the ejecta spectra (see \S \ref{subsect-pl+ej}) using free
neutral hydrogen column density.  We found column density values
roughly consistent with those quoted above for the featureless
rims. Again, we find a slightly larger absorption in the W
($\mathrm{N}_{\mathrm{H}} = 0.72 \pm 0.02 \times 10^{22} \:
\mathrm{cm}^{-2}$) than in the NW ($0.66 \pm 0.02 \times 10^{22} \:
\mathrm{cm}^{-2}$) and NE ($0.61 \pm 0.02 \times 10^{22} \:
\mathrm{cm}^{-2}$). Considering these $\mathrm{N}_{\mathrm{H}}$ values
as well as the ones derived from the simple power-law fits (see
above), we chose to fix the column density to an approximate mid-point
value of $\mathrm{N}_{\mathrm{H}}= 0.7 \times 10^{22} \:
\mathrm{cm}^{-2}$ for the subsequent analyses.

These pure power-law fits are clearly incomplete in that they do not
account for obvious emission lines in the spectra. The presence of
these lines argues for a thermal component that varies across the
radial sequence. Further there is the possibility that the observed
steepening in the featureless spectra may be due to this additional
thermal component, which, being softer than the power-law component
and growing in contribution moving inward, causes an apparent
softening of the spectral index. To test this possibility, we study
two specific situations where the thermal emission comes either from
the shocked ejecta (\S \ref{subsect-pl+ej}) or from the shocked
ambient medium (\S \ref{subsect-pl+nei}).

\subsubsection{Power-law + ejecta template \label{subsect-pl+ej}}

To investigate how the observed steepening may be modified by the
introduction of a shocked ejecta thermal component, we take for each
sector (i.e., W, NW, and NE) a template model spectrum of the shocked
ejecta and see how much of it can be included along with a power-law
in the various radial regions. Each azimuthal region has a different
template spectrum determined by fitting the data extracted from nearby
inner regions of the remnant (see label ``Ejecta'' in Table
\ref{tab-regions}) to single-component and constant-temperature
non-equilibrium ionization (NEI) spectral models.

\begin{figure*}[t]
\centering
\includegraphics[bb=0 35 566 850,clip,width=13cm]{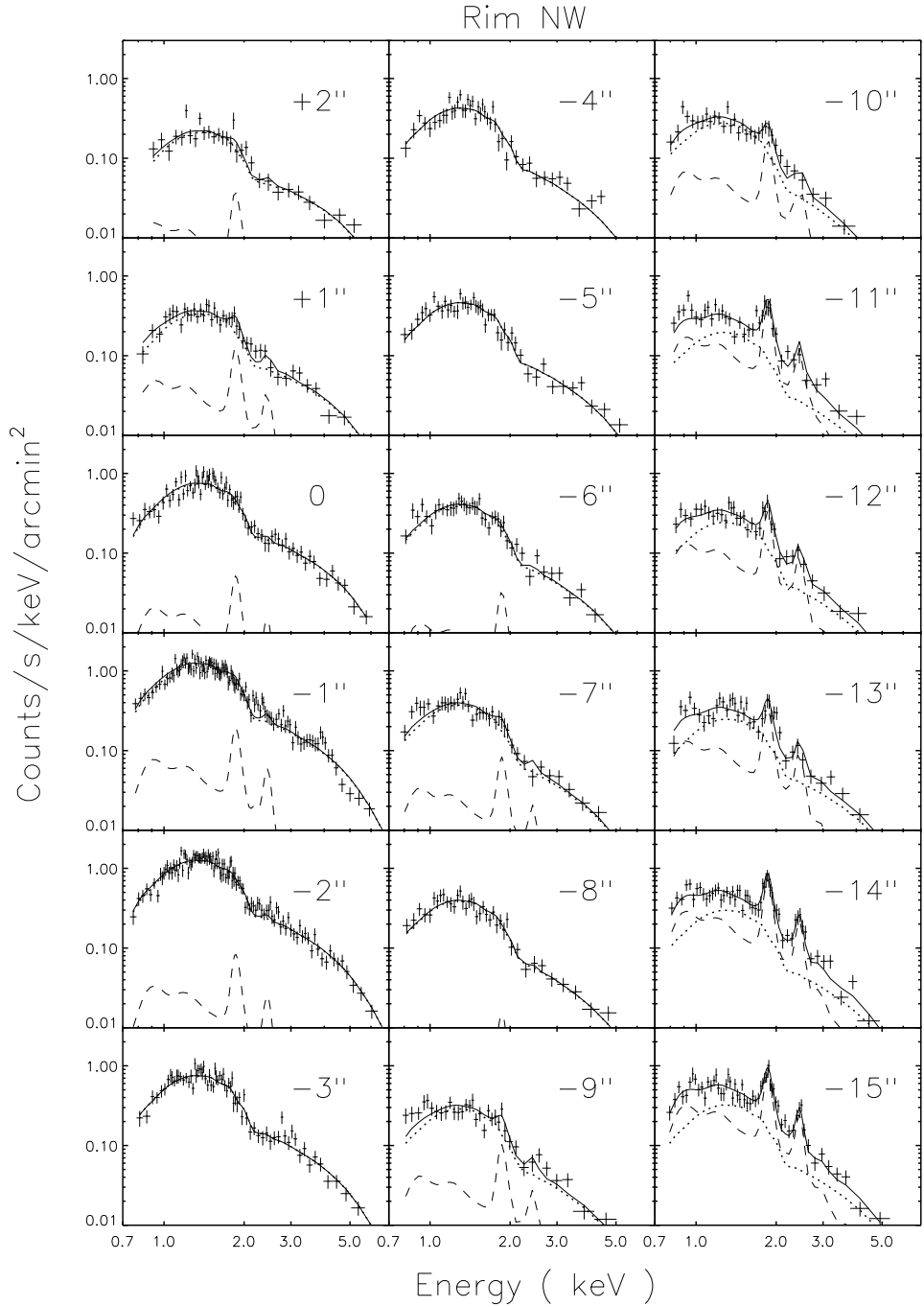}
\caption{Spectra extracted from the rim NW (see Fig.
\ref{fig_Tycho}) fitted with a power-law model (dotted line) plus
a template for the shocked ejecta (dashed line), the absorption
being held fixed to $0.7 \times 10^{22} \: \mathrm{cm}^{-2}$. The
sum of the two models is shown in solid line. The numbers give the
position from the blast wave in arcseconds.}
\label{fig_rim_spectra_grid_rimNW_PLwEJ}
\end{figure*}

\begin{figure*}[t]
\centering
\includegraphics[bb=0 35 566 850,clip,width=13cm]{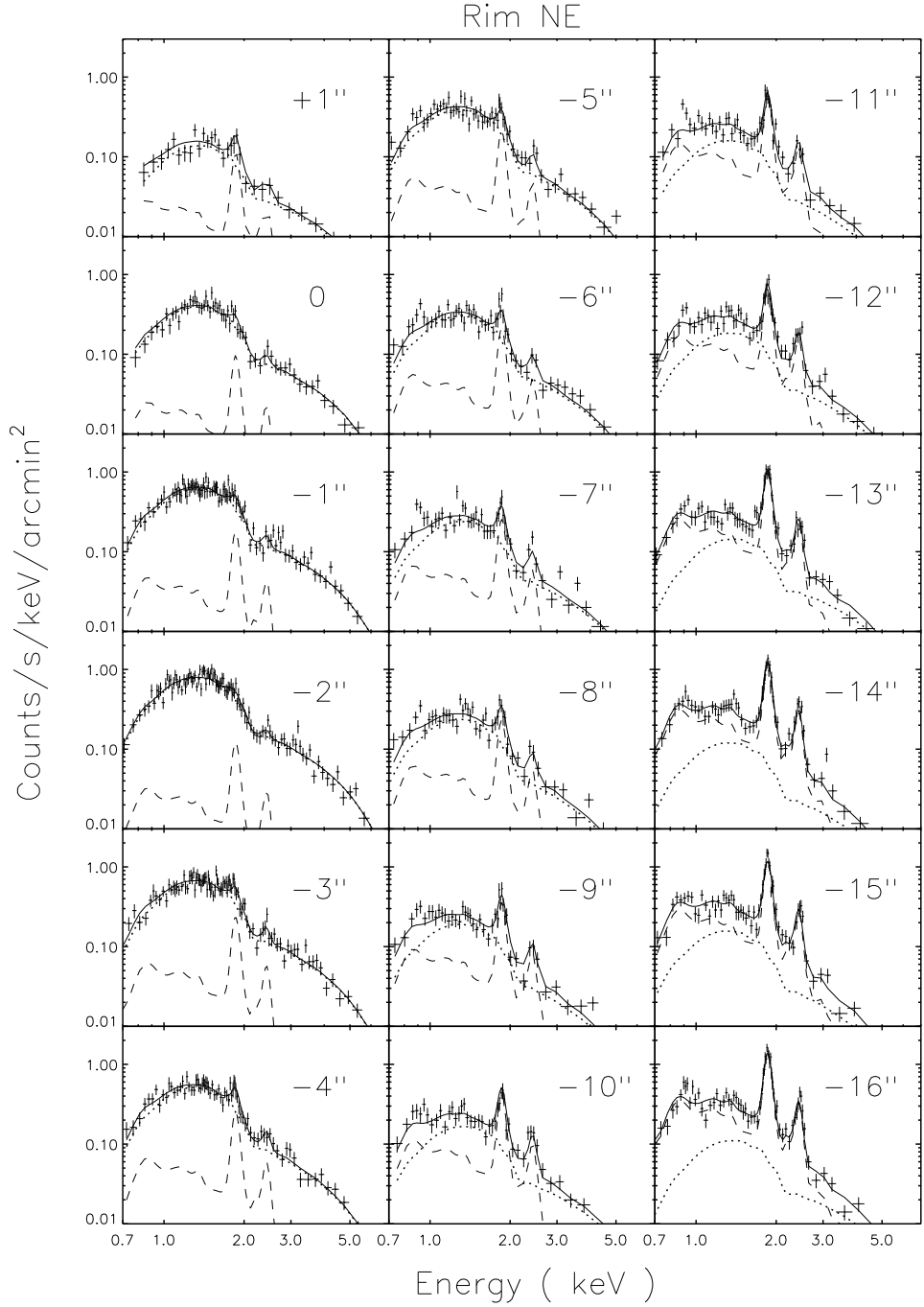}
\caption{Same as Fig. \ref{fig_rim_spectra_grid_rimNW_PLwEJ} but
for the rim NE (see Fig. \ref{fig_Tycho}).}
\label{fig_rim_spectra_grid_rimNE_PLwEJ}
\end{figure*}

Figure \ref{fig_param_vs_radius_mo_all_one_fig} shows the best-fit
parameters derived from a power-law model to which we add an ejecta
component ({\Large $\star$} data points), with the absorption held
fixed, in the W, NW and NE rims. We can see how the profile of the
photon index (top panel) is modified by comparing to the single
power-law model fits ({\Large $\bullet$} data points). The spatial
variations of the photon index profile are unchanged behind the blast
wave, although there is some modification in the inner regions
dominated by strong emission lines (see Table
\ref{tab-slope-xspec-fit} for the rim W). As shown by the reduced
$\chi^2$ profile (bottom panel), the introduction of the ejecta
template improves the fit quality in the regions where emission lines
are strong as we would have expected, but sometimes also in regions
where the emission lines are faint (see rims W and NE). This latter
point strongly suggests that small knots of shocked ejecta have nearly
reached the blast wave where they contribute significantly to the
X-ray emission.  Figures \ref{fig_rim_spectra_grid_rimNW_PLwEJ} and
\ref{fig_rim_spectra_grid_rimNE_PLwEJ} illustrate the respective
contribution of the shocked ejecta and power-law components in the NW
and NE regions.

We note, however, that the radial regions where the $\chi^2$
difference between models with and without the ejecta template becomes
important (bottom panel of
Fig.~\ref{fig_param_vs_radius_mo_all_one_fig}), found at about
$-9\arcsec$ in the rims W and NW and $-4\arcsec$ in the rim NE with
respect to the blast wave, do not correspond to the mean position of
the contact discontinuity position. At these radii the featureless
emission from the forward shock is still the dominant broadband
spectral component. Rather, these locations likely represent the
outermost ``fingers'' of shocked ejecta resulting from the
Rayleigh-Taylor instability at the contact discontinuity.  On average,
the contact discontinuity lies much further in (see \S
\ref{subsect-BW-CD}).

\subsubsection{Power-law + NEI model \label{subsect-pl+nei}}

We consider now that the line emission could be entirely due to the
shocked ambient gas instead of considering, as we did in the previous
section, that such emission is associated with the ejecta
material. Therefore we introduce, in addition to the power-law, a
thermal component associated with the shocked ambient medium and see
how this component affects the steepening of the power-law model found
in our previous analysis.

The thermal model that we choose for this spectral analysis is a
simple NEI model with solar abundances  \citep[here][]{ang89}.
The parameters of this model are the electronic temperature
$kT_{\mathrm{e}}$, the ionization age $\tau \equiv \int_{t_s}^{t_0} \:
n_{\mathrm{e}}(t) \: dt$ where $n_{\mathrm{e}}$ is the post-shock
electronic density and $t_0-t_s$ is the flow time (i.e., the time
since shock-heating), and the emission measure $E_{\mathrm{X}} \propto
n_{\mathrm{e}}^2 \: V / D^{2}$ where $V$ is the emission volume and
$D$ the distance to the remnant.

When fitting the data with a power-law and the above thermal model
(the absorption being fixed), spectral fits (not shown here) were, in
the radial regions where emission lines are still faint, as good as or
sometimes even slightly better than those obtained using the simple
power-law model with fixed absorption. However, the model did not
place the lines at the observed positions in the spectra of the
quasi-featureless regions. In addition, there was no consistent
pattern in the radial variation of the ionization age, while the
ionization age of the inner regions was far too low to be consistent
with the density required to explain the observed brightness.

Considering this last point, we introduce a new NEI model for the
shocked ambient medium where both the ionization age and emission
measure are consistent with the same post-shock electronic density. We
refer to this as the ``self-consistent plane shock NEI model'' or
SCPNEI for short. To use this model, we need to determine the flow
time and the volume of each emitting region, which in turn gives the
ionization age and the emission measure, assuming a given value for
the post-shock electronic density. To estimate the flow time, we used
a semi-analytical hydrodynamical model (to be described in more detail
in \S \ref{subsect-cr-hydro-model} below) that is capable of
reproducing the observed ratio of the blast wave to contact
discontinuity radii. We use the same flow time over the entire volume along the line of sight at a given projected distance from the shock (this is of course imperfect, the true flow time is shorter close to the shock in the front and in the back). The volumes are determined numerically. Each $1\arcsec$-wide region has then a single-component
NEI model with its own ionization age (fixed by the post-shock
electronic density) and in which the electronic temperature is the
only fitting parameter.

In the following, we restrict the analysis to rim W which has the
highest statistics and where the contact discontinuity is the farthest
from the blast wave. Volumes and flow times at each radial position
behind the blast wave (i.e., $0, -1\arcsec, \ldots, -15\arcsec$) are
given in Table \ref{tab-nei-hydro+ej-xspec-fit} (see the three first
columns).  For each position, we looked at the variations of the
photon index, electronic temperature and $\chi^2$ of our power-law
plus SCPNEI model as a function of the post-shock electronic density.
While we varied the post-shock electronic density over more than two
orders of magnitude starting from $0.1 \: \mathrm{cm}^{-3}$, we found
the power-law index profile to be very stable in the regions of lowest
$\chi^2$. The largest differences in the spectral index were about 0.1
which is far too small to affect the observed steepening.

\subsubsection{Power-law + NEI model + ejecta template\label{subsect-pl+nei+ej}}

\begin{figure*}[t]
\centering
\includegraphics[width=7cm]{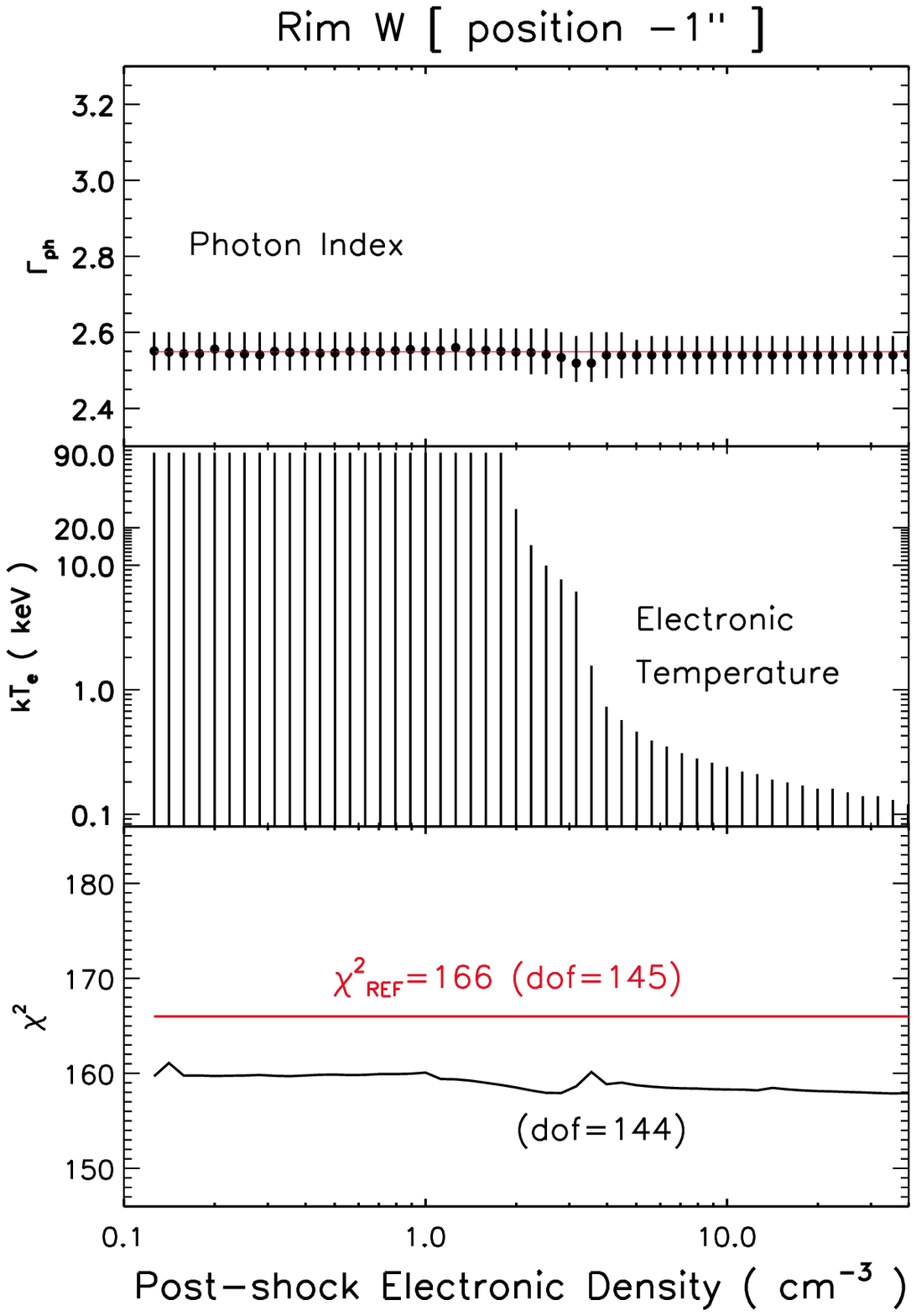}
\includegraphics[width=7cm]{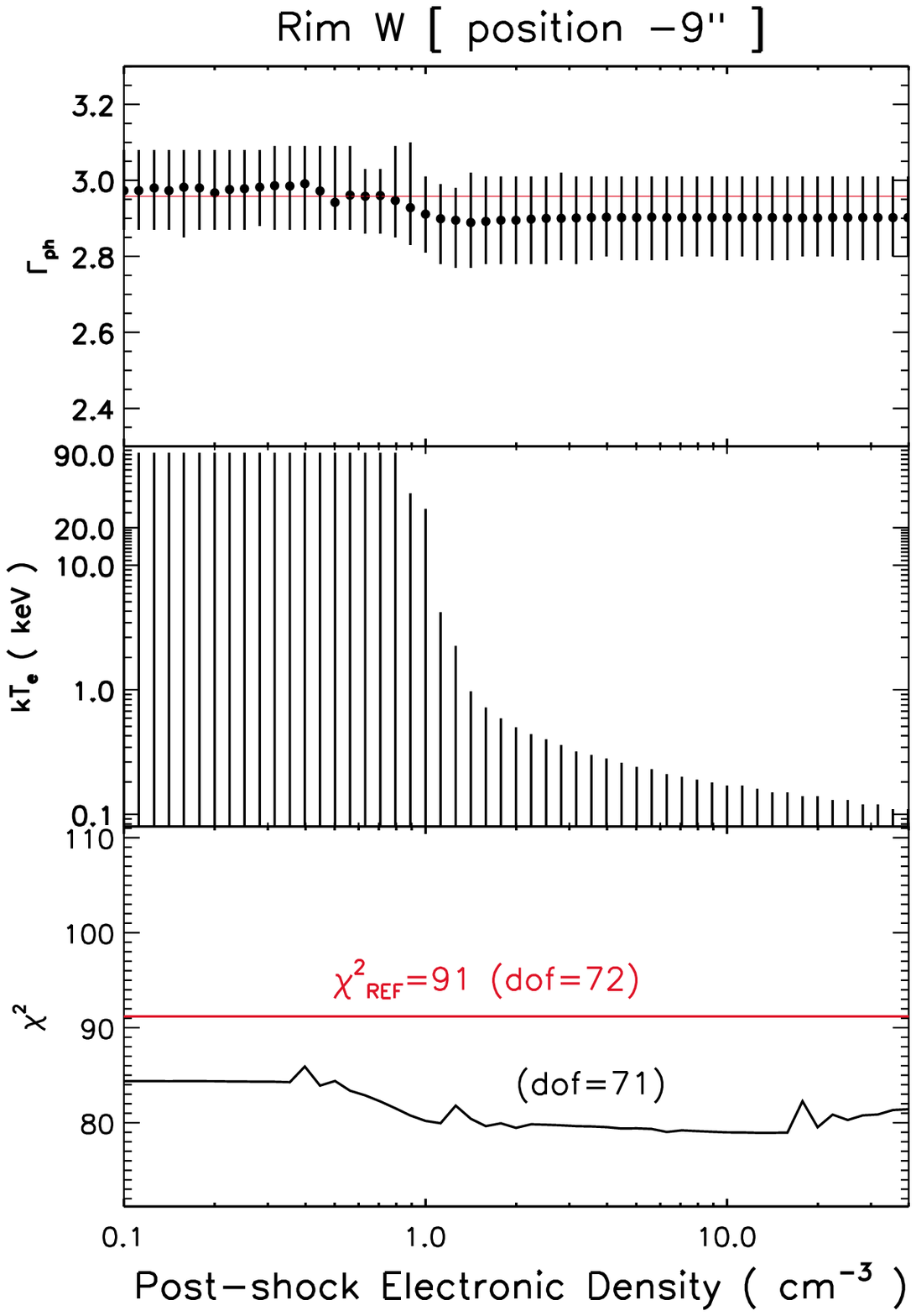}
\caption{Photon index $\Gamma_{\mathrm{ph}}$ (top panels), electronic
  temperature $kT_{\mathrm{e}}$ (middle panels) and best-fit $\chi^2$
  values (bottom panels) obtained by fitting the spectra of the rim W
  (see Fig. \ref{fig_rim_spectra_grid_rimW}) with a power-law model
  plus a template for the shocked ejecta and a NEI model where both
  the ionization age and emission measure are linked (see \S
  \ref{subsect-pl+nei+ej}) as a function of the post-shock electronic
  density. This is shown for two regions behind the blast wave (left:
  $-1\arcsec$ and right: $-9\arcsec$).  The red lines show the values
  obtained with only a power-law model plus an ejecta template
  ($\chi^2 = \chi^2_{\mathrm{REF}}$).  The error bars plotted
  correspond to the range $\Delta \chi^2 \equiv \chi^2_{\mathrm{REF}}
  - \chi^2 < 0$.  In these models, the absorption was held fixed to
  $0.7 \times 10^{22} \: \mathrm{cm}^{-2}$.  The rise in $\chi^2$ at
  high density in the $-9\arcsec$ panel is due to the lack of low
  temperature models in XSPEC.}
\label{fig_data_REG_waponei_link_tau_norm_nh0.7}
\end{figure*}

Because we showed in \S \ref{subsect-pl+ej} that introducing an ejecta
template improves the fit very significantly, the previous procedure
(\S \ref{subsect-pl+nei}) was repeated by introducing the ejecta
template along with the power-law and SCPNEI model (see two examples
in Fig.~\ref{fig_data_REG_waponei_link_tau_norm_nh0.7}).  For this
ultimate case, very similar results were obtained in terms of
stability of the power-law index steepening (top panel of
Fig.~\ref{fig_data_REG_waponei_link_tau_norm_nh0.7}). With the
introduction of the SCPNEI model, the spectral fits are slightly
improved compared to a power-law model plus ejecta template (bottom
panel of Fig.~\ref{fig_data_REG_waponei_link_tau_norm_nh0.7}).  This
is however only marginally significant (at most $\Delta \chi^2 \equiv
\chi^2_{\mathrm{REF}} - \chi^2= -12$ for the inclusion of two
additional parameters, where $\chi^2_{\mathrm{REF}}$ is the $\chi^2$
of the power-law plus ejecta template model).

This method allows us, in addition, to constrain the electronic
temperature as a function of the post-shock electronic density (middle
panel of Fig.\ \ref{fig_data_REG_waponei_link_tau_norm_nh0.7}).
Because we do not convincingly detect the shocked ambient medium, we
use the power-law plus ejecta model as a reference and define the
allowed domain as $\Delta \chi^2 < 0$ (i.e., we require that the
additional SCPNEI component not degrade the fit). As expected, when
the electronic density is low (below $\sim 2 \: \mathrm{cm}^{-3}$ at
position $-1\arcsec$ and $\sim 1 \; \mathrm{cm}^{-3}$ at position
$-9\arcsec$), the range of allowed temperatures is not constrained but
as the density gets larger, a forbidden high-temperature regime
appears (above a few keV at position $-1\arcsec$ and above 1 keV at
position $-9\arcsec$).

\subsection{Where is the shocked ambient medium?\label{subsect-sam}}

The previous analysis has presented solid evidence for a spectral
index variation in the synchrotron emission behind the blast wave in
Tycho. We have also presented evidence for significant X-ray thermal
emission from shocked ejecta far ahead of the contact discontinuity in
the blast wave zone. However, evidence for a shocked ambient medium
thermal component remains weak. The data are compatible with no such
thermal component and provide constraints on its density and
temperature. We found, in particular, that a large range of post-shock
electronic densities were allowed when fitting the spectra at each
individual position behind the blast wave with poor constraints on the
electronic temperature for low densities. Even rather large density
values were allowed as long as the electronic temperature was quite
low.  However there are constraints on the allowed post-shock
temperatures that come from the hydrodynamic evolution of the remnant,
as we utilize here.

\subsubsection{CR-hydro NEI model\label{subsect-cr-hydro-model}}

To better constrain the pre-shock ambient density, we use
self-consistent temperature and hydrodynamical profiles whose
parameters are determined from a semi-analytical model of cosmic-ray
modified SNR hydrodynamics which borrows a 1-D similarity solution for
the hydrodynamic variables. Our simulations based on this self-similar
hydrodynamical calculation are coupled with a nonlinear diffusive
shock acceleration model, so that the backreaction of the particles
accelerated at the blast wave is taken into account
\citep[see][]{dee00}. This is important because these simulations,
unlike those that treat the accelerated particles as test-particles,
are able to reproduce the observed blast wave to contact discontinuity
radii ratio of $\sim 1.11$ that we found in rim W (see \S
\ref{subsect-BW-CD}). This constraint results in an overall
compression ratio, $r_{\mathrm{tot}}$, close to 6.

\begin{figure*}[t]
\centering
\includegraphics[bb=0 20 360 506,clip,width=11cm]{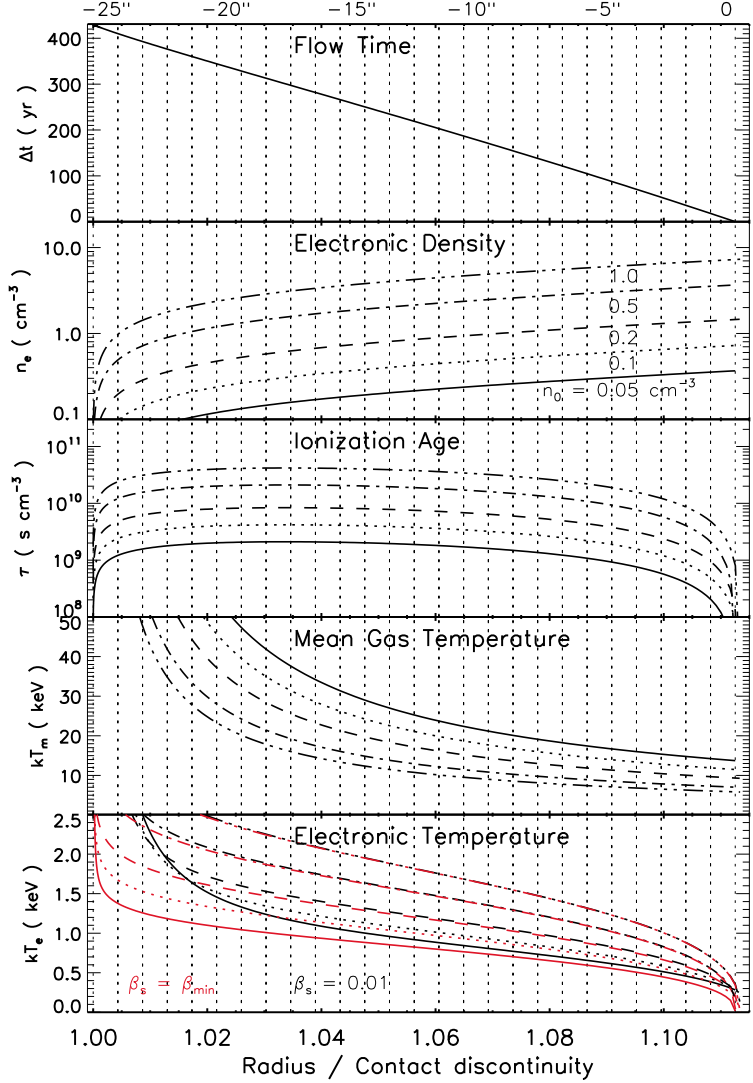}
\caption{Parameters obtained from models of SNR evolution that include
  the backreaction from accelerated particles. The ratio of the blast
  wave to the contact discontinuity radii matches the one observed in
  the rim W (i.e., $R_s/R_c = 1.113$). Radial profiles of the flow
  time $\Delta t$, electronic density $n_{\mathrm{e}}$, ionization age
  $\tau$, mean gas temperature $kT_{\mathrm{m}}$, electronic
  temperature $kT_{\mathrm{e}}$ are shown for different pre-shock
  ambient medium densities: $n_0 = 0.05$ (solid line), $0.1$ (dotted
  line), $0.2$ (dashed line), $0.5$ (dash dot line) and $1.0$ (dash
  dot dot line). The electronic temperature profile was computed
  assuming an initial electron-to-proton temperature ratio
  $\beta_{\mathrm{s}}$ at the blast wave of
  $\beta_{\mathrm{s}}=\beta_{\mathrm{min}}$ (red lines) where
  $\beta_{\mathrm{min}}$ is typically the electron-to-proton mass
  ratio (i.e., zero-equilibration) and 0.01 (black lines). The case of
  full temperature equilibration ($\beta_{\mathrm{s}} = 1$) is
  directly given by the mean shock temperature profile. The radial
  profiles between the blast wave and contact discontinuity were split
  up into a number of shells (vertical dotted lines) equal to the one
  in the observation and the average values of $n_{\mathrm{e}}$,
  $\tau$ and $kT_{\mathrm{e}}$ within each shell will serve as input
  for a thermal model (see \S
  \ref{subsect-cr-hydro-model}).} \label{fig_nei_param_alln0}
\end{figure*}

For a given ambient medium density, the hydrodynamic model provides
radial profiles of the flow time, electronic density, ionization age
and mean shock temperature from the blast wave to the contact
discontinuity (Fig.\ \ref{fig_nei_param_alln0}).  Assuming a given
electron-to-proton temperature ratio at the blast wave, the electron
temperature profile resulting from Coulomb collisional heating can be
calculated from the mean shocked gas temperature and density profiles
\citep{it77, coa82}. The self-consistent NEI model that we introduce
and will use here (hereafter the ``CR-hydro NEI'' model) is based on
these profiles, which we divided into several shells to match the
number of zones in rim W (see Fig. \ref{fig_nei_param_alln0}). Each
shell can be characterized by a set of average spectral parameters.

Because of projection effects, the NEI model of a given observed zone
(i.e., here a $1\arcsec$-wide sector region) is not the NEI model of
the corresponding simulated shell.  Indeed, each zone seen in
projection onto the sky includes the contribution from a specific
number of shells. Our CR-hydro NEI model in one zone is then the
combination of several single-component constant-temperature NEI model
from the shells.  The volume contributions of the different shells to
a given projected zone are computed numerically.  These volumes and
the post-shock electronic density, together with the distance
(corresponding to the simulated blast wave radius, see Table
\ref{tab-CR+hydro-simu}) allow us to derive the emission measure of
each single NEI model.

\subsubsection{Initial parameters of the CR-hydro model\label{subsect-initial-param}}

Cosmic-ray modified hydrodynamics models were run for different values
of the ambient medium density (see Table
\ref{tab-CR+hydro-simu}). There are two different ways to match the
observed radii ratio between the blast wave and contact discontinuity
in the model. We could vary either the injection efficiency,
$\eta_{\mathrm{inj}}$, which is the fraction of total particles which
end up with suprathermal energies, or the unshocked upstream magnetic
field, $B_0$ \citep{bee99, elb00}. An increase in the injection
efficiency increases the cosmic-ray pressure and then the overall
compression ratio, while an increase in the upstream magnetic field
increases the heating of the gas by the Alfv\'en waves in the
precursor region and then tends to reduce the overall compression
ratio \citep[see][]{bee99}. Since the closeness of the blast wave and
contact discontinuity (radii ratio of $\sim 1.1$) is suggestive of
efficient particle acceleration, we choose to set the injection
parameter ($\eta_{\mathrm{inj}}$) to a high value of $10^{-3}$ and
adjust the upstream magnetic field for each run (see Table
\ref{tab-CR+hydro-simu}). This case leads to lower post-shock gas
temperatures than those derived from models with lower injection
efficiency and hence more conservative limits on the value of the
ambient density. Increasing $\eta_{\mathrm{inj}}$ to an even higher
value of $10^{-2}$ would in fact reduce further the post-shock
pressure of the thermal particles albeit only by something like 15\%,
which would not significantly increase our density limits.

We fixed the ejected mass and kinetic energy of the ejecta to $1.4 \;
\mathrm{M}_{\odot}$ and $E_{\mathrm{SN}} = 10^{51} \: \mathrm{ergs}$,
respectively, which are standard values for thermonuclear SNe. The age
of the SNR was fixed to $430$ years, approximately the age of
Tycho. We choose a power-law index of the initial power-law density
profile in the ejecta, $n$, equal to 7 so that the expansion parameter
in the model ($m = 1-3/n = 4/7\simeq 0.57$) is consistent with the
mean expansion parameter derived from the X-ray observations
\citep[$0.54\pm0.05$,][]{hu00}.  A slightly higher index $n$ would
produce a narrower gap between the blast wave and contact
discontinuity (in the test-particle limit, we have for instance
$R_s/R_c = 1.14$ for $n=9$ compared to $R_s/R_c = 1.18$ for $n=7$) but
would clearly overestimate the remnant's expansion rate ($m \simeq
0.67$ for $n=9$) compared to the observations. Below, we discuss the
difference between the use of a power-law and exponential
distributions for the initial density profile in the ejecta (see \S
\ref{subsect-behind-rim}).  We assume that the SNR evolves into an
interstellar medium that is uniform in density and magnetic field and
whose pressure is $2300 \: \mathrm{K} \: \mathrm{cm}^{-3}$.

The self-similar model is in principle not valid as soon as the
reverse shock reaches the core (flat density profile) of the
ejecta. For $n=7$ and $M_{\mathrm{ej}} = 1.4 \; \mathrm{M}_{\odot}$,
the mass in the ramp is $\frac{3}{n} \: M_{\mathrm{ej}} = 0.6 \;
\mathrm{M}_{\odot}$. The mass swept-up by the blast wave,
$M_{\mathrm{sw}}$, when the reverse shock reaches that point is twice
that for $n=7$ (for no CRs), so the model is good for $M_{\mathrm{sw}}
< 1.2 \; \mathrm{M}_{\odot}$. This is reached somewhere between $n_{0}
= 0.1 \: \mathrm{cm}^{-3}$ and $0.2 \: \mathrm{cm}^{-3}$ in Table
\ref{tab-CR+hydro-simu}. This means that the self-similar model does
not apply to higher densities. In particular, to get the same blast
wave to contact discontinuity radii ratio when the reverse shock is
inside the ejecta core already would presumably require a larger
compression ratio to begin with. Addressing that goes beyond the scope
of this paper and would require use of a numerical hydrodynamical
code.

\subsubsection{Constraints on the ambient medium density\label{subsect-constraint-n0}}

\begin{figure}[t]
\centering
\includegraphics[width=8cm]{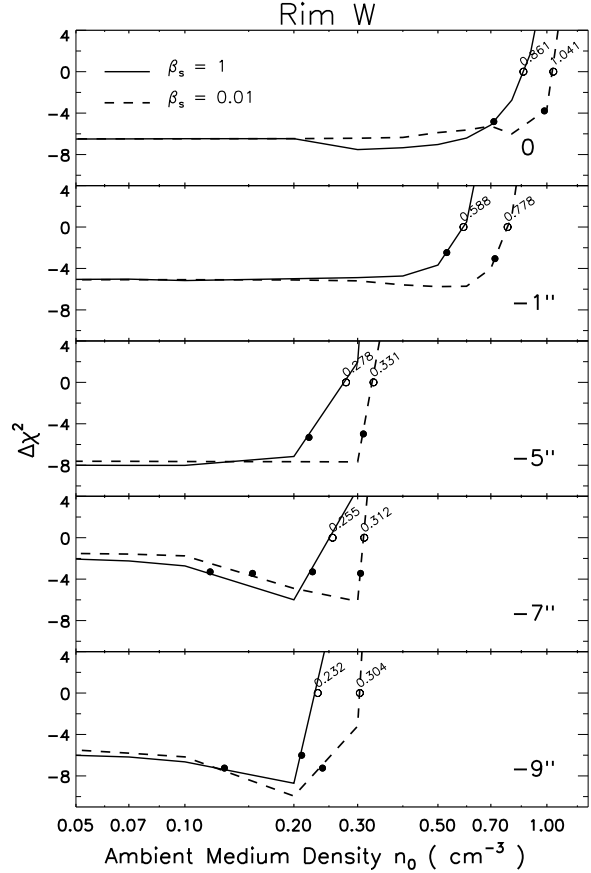}
\caption{Difference between the $\chi^2$ obtained from a power-law
+ ejecta template + CR-hydro NEI model and the one obtained from a
power-law + ejecta template (labelled as $\chi^2_{\mathrm{REF}}$
in Table \ref{tab-slope-xspec-fit}) as a function of the pre-shock
ambient medium density $n_0$, for different electron-to-proton
shock temperature ratios: $\beta_{\mathrm{s}} = 1$ (solid line)
and $0.01$ (dashed line). This is shown for different regions
behind the blast wave ($0, -1\arcsec, -5\arcsec, -7\arcsec,
-9\arcsec$) in the rim W. The filled circles ({\Large $\bullet$})
correspond to points where $\Delta \chi^2=\min \Delta \chi^2 +
2.7$ and the open circles ({\Large $\circ$}) to points where
$\Delta \chi^2=0$. The corresponding densities are labelled for
this latter case.} \label{fig_plot_chi_vs_n0_beta}
\end{figure}

Figure \ref{fig_plot_chi_vs_n0_beta} shows the results obtained
from the modelling and spectral fitting of several radial regions
in rim W. We plot the $\chi^2$ difference between a power-law +
ejecta template + our CR-hydro NEI model and a power-law + ejecta
template as a function of the ambient medium density. We start
from the blast wave (top panel, position $0$) to the inner regions
(bottom panel, position $-9\arcsec$). The different curves
correspond to different initial values of the electron-to-proton
temperature ratio at the blast wave ($\beta_{\mathrm{s}}=1$ in
solid lines and $\beta_{\mathrm{s}}=0.01$ in dashed lines). Note
that the introduction of the CR-hydro NEI model whose parameters
are all fixed will not necessarily always improve the fit. We only
show a few regions in Figure \ref{fig_plot_chi_vs_n0_beta} where
the $\Delta \chi^2$ curve goes negative below a certain density.

The points where the $\chi^2$ difference is null correspond to
strong upper limits on the ambient medium density $n_0$. For a
given electron-to-proton temperature ratio $\beta_{\mathrm{s}}$ at
the shock, these upper limits are more and more refined as we
probe the innermost regions. The most stringent limits obtained at
position $-9\arcsec$ are $n_0 \lesssim 0.2 \: \mathrm{cm}^{-3}$
for $\beta_{\mathrm{s}}=1$ and $n_0 \lesssim 0.3 \:
\mathrm{cm}^{-3}$ for $\beta_{\mathrm{s}}=0.01$. We note that the
case $\beta_{\mathrm{s}}=0.01$ is not very different from the case
of minimum equilibration $\beta_{\mathrm{s}}=\beta_{\mathrm{min}}$
(where $\beta_{\mathrm{min}}$ is the minimum electron-to-proton
temperature ratio at the shock given by the electron-to-proton
mass ratio) for low ambient medium densities (below $\sim 1 \:
\mathrm{cm}^{-3}$) as long as we stay near the blast wave (i.e.,
roughly between position 0 and position $-15\arcsec$ in Fig.
\ref{fig_nei_param_alln0}).

\begin{figure*}[t]
\centering
\includegraphics[bb=0 35 566 850,clip,width=13cm]{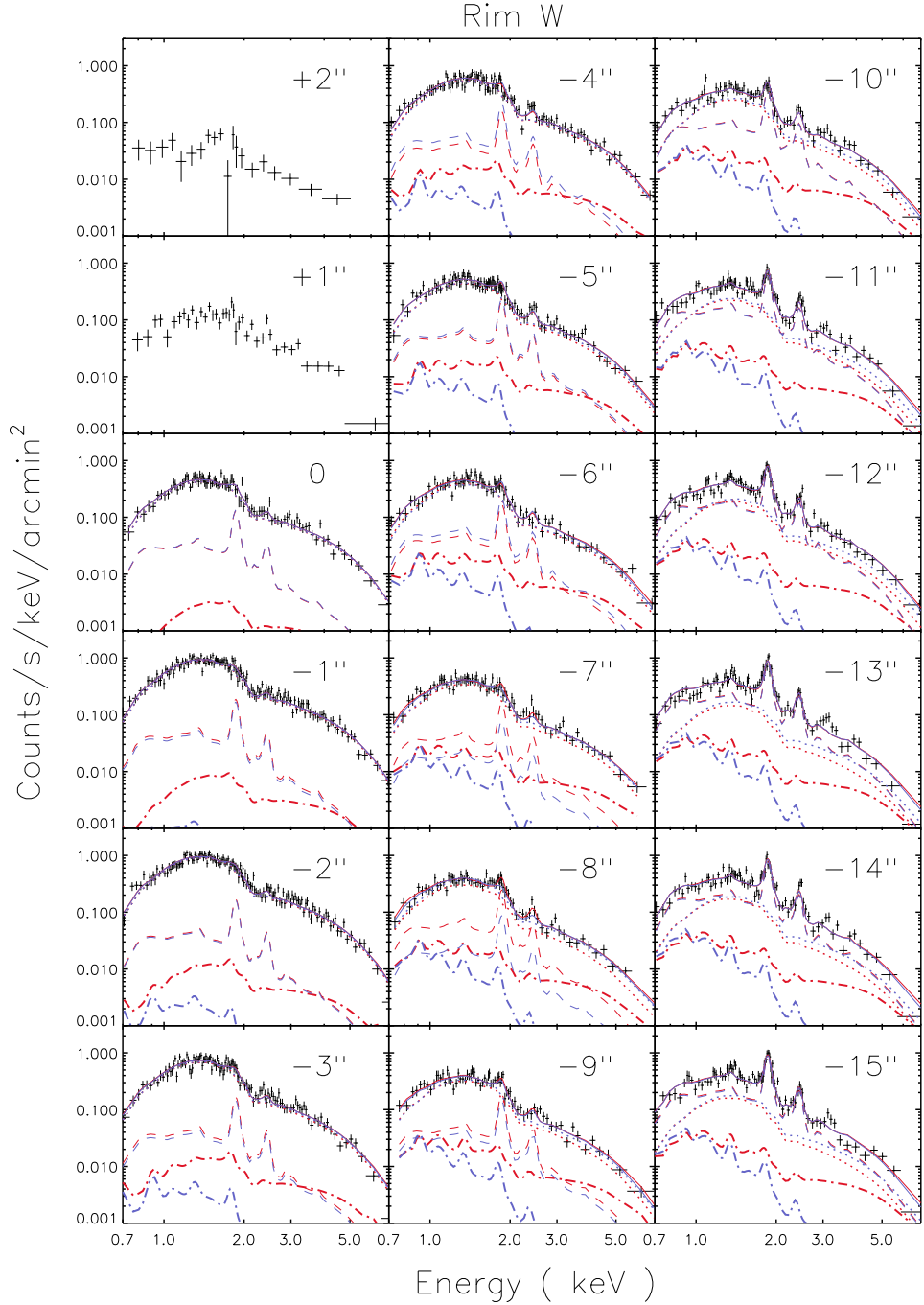}
\caption{Different components of a best-fit model (to spectra
  extracted from the W rim) that includes a power-law (dotted line), a
  shocked ejecta template (dashed line) and a self-consistent NEI
  model (thick dash dot line), the interstellar absorption being held
  fixed to $0.7 \times 10^{22}$ cm$^{-2}$. The parameters of the NEI
  model are derived from a CR-hydro model and take into account
  projection effects. The sum is shown in solid line. We show two
  cases that correspond to the same pre-shock ambient medium density
  of $n_{0} = 0.2 \: \mathrm{cm}^{-3}$ but to different
  electron-to-proton temperature ratios at the shock:
  $\beta_{\mathrm{s}} = 1$ (red) and 0.01
  (blue).} \label{fig_rim_spectra_grid_rimW_PLwEJwNEI_hydro_n0}
\end{figure*}

In Figure \ref{fig_rim_spectra_grid_rimW_PLwEJwNEI_hydro_n0}, we
illustrate for the best-fit model the respective contributions of the
power-law, shocked ejecta and shocked ambient medium for two
extreme values of the electron-to-proton temperature ratio at the
shock ($\beta_{\mathrm{s}} = 0.01$ and $\beta_{\mathrm{s}} = 1$)
and an ambient density of $n_0 = 0.2 \: \mathrm{cm}^{-3}$ favored
by the X-ray data. Table \ref{tab-nei-hydro+ej-xspec-fit} gives
the associated photon index values.

Our conclusion of this spectral analysis is that the ambient
medium density must be less than $0.3 \: \mathrm{cm}^{-3}$. This
corresponds to a lower limit on the distance to the remnant of
$\sim 2.8 \; \mathrm{kpc}$ (Table \ref{tab-CR+hydro-simu}) which
is consistent with the upper limit range derived from optical
observations \citep{chk80, smk91} and in good agreement with the
value derived from analysis of the historical light curve
\citep{ru04}.

\subsection{Radio and X-ray radial profiles\label{subsect-radio-xray}}

\subsubsection{Radio to X-ray comparison\label{subsect-radio-xray-compare}}

The first observational test to determine whether the nonthermal
X-ray rims are limited by the magnetic field or by the energy
losses of the radiating electrons consisted of searching for
radial variations of the X-ray synchrotron spectrum and
particularly for radial variations of its slope (see \S
\ref{subsect-PLindex}). A second observational test, which also
allows us to gain some insight into the spatial distribution of
the magnetic field and accelerated electrons, consists of
comparing radio and X-ray maps of the non-thermal emission at the
rim (see \S \ref{subsect-at-rim}). If the X-ray rims are limited
by the synchrotron losses, the radio synchrotron emission should
be much broader than the X-ray synchrotron emission because the
radio-emitting electrons are not affected by radiative losses. If
the X-ray rims are magnetically limited, one expects radio rims as
well, but wider and with a smaller brightness contrast between the
rim and far behind the rim \citep{poy05}.

Figure \ref{fig_radio-xray_profiles} shows the radio and X-ray radial
profiles of rims W, NW and NE. The X-ray radial profiles (data points
marked with {\Large $\bullet$}, scale on the left) were obtained from
the 2003 \textit{Chandra} image in the 4-6 keV continuum energy bands
which emphasize the thin rims at the blast wave. The radio profiles
(solid lines, scale on the right) were obtained from the 1995
\textit{VLA} image and were shifted by $3\arcsec$ to compensate for
the remnant's expansion.  The same spatial regions were averaged in
the radio image, using the X-ray-derived radii of curvature (Table
\ref{tab-regions}).  In this comparison, we implicitly assume that the
radio flux did not change over 8 years. We note that the radio
profiles from the NW and NE rims are similar in peak intensity, while
the W rim is about a factor of three fainter. The comparison between
the radio and the X-ray profiles clearly shows that the X-ray rims
have a radio counterpart characterized by a larger width and a smaller
brightness contrast between the rim and the center in the NW and NE
rims (but not in the W rim). This does not imply however that the
X-ray rims are magnetically limited because the most critical
constraints are the absolute radio flux and the ratio of X-ray to
radio fluxes (see \S \ref{subsect-synchproj-profile}).

\subsubsection{Confidence in the radio
data\label{subsect-radio-xray-confidence}}

There is a contradiction between the radio profile presented in this
paper (see Fig. \ref{fig_radio-xray_profiles}), which were based on
the data taken in 1994-1995 and presented by \cite{rem97}, and the one
obtained by \cite{div91} in the late 1980's (see their Fig.~2). The
profile presented by \cite{div91}, a slice across the remnant that
goes from S-SW to N-NE, shows a clear highly peaked outer filament on
the N-NE side with a factor of four drop from this peak to the next
valley. On the other hand, our attempt\footnote{Note that the
  declination value quoted in the caption to figure 2 of \cite{div91}
  is not given in proper sexigesimal notation and therefore may be in
  error.}  to reproduce the profile of \cite{div91} with the radio
data of \cite{rem97} shows at best a drop of only a factor of
two. This difference is larger than what we would expect to obtain by
changing the method to reconstruct the radio image. We did verify that
profiles extracted from the radio data we have in hand match closely
the profiles published by \cite{rem97}.

This difference suggests a possible problem in either the radio
data of \cite{div91} or those of \cite{rem97}. In this paper, we
chose to use the data presented by \cite{rem97} because they are
the most recent. In addition, in the following analysis, because
of the above uncertainties, we will try not to use local flux
values extracted from the radio image as inputs for models. A
careful study of the relative radio and X-ray nonthermal emissions
needs better, and better characterized, radio data.

%*******************************************
\section{Discussion\label{sect-discussion}}
%*******************************************

The two astrophysical questions that we address in this paper can
be simply expressed as follows: (1) why is the X-ray emission so
bright at the blast wave and (2) why does the X-ray emission fall
so rapidly to faint values behind the bright rim?

The first question aims to understand the origin of the rim
morphology. Is the bright X-ray synchrotron emission caused by a
magnetic field locally very high at the blast wave or by the fact
that the highest energy electrons cannot travel far from their
acceleration site without losing energy so that their emission is
concentrated to very thin regions just behind the blast wave (\S
\ref{subsect-at-rim})?

The second question aims to understand the origin of the absence
of thermal X-ray emission from the shocked ambient medium. Is this
absence caused by a low ambient medium density so that the thermal
X-ray emission is overwhelmed by the X-ray synchrotron emission or
by a low temperature plasma presumably resulting from efficient
particle acceleration so that the thermal emission is hidden by
interstellar absorption or shifted below the X-ray domain (\S
\ref{subsect-behind-rim})?

\subsection{The dark side of the rim\label{subsect-behind-rim}}

In light of the results we present in \S \ref{subsect-sam}, we
conclude that the lack of thermal X-ray emission from the shocked
ambient medium between the blast wave and the contact discontinuity
essentially reflects a low density in the ambient medium around Tycho.

At the blast wave, we found that the most stringent upper limit on the
ambient medium density is about $0.9 \: \mathrm{cm}^{-3}$ if electrons
and protons are in temperature equilibrium and $1 \: \mathrm{cm}^{-3}$
if there is no equilibrium (\S \ref{subsect-constraint-n0} and top
panel of Fig.  \ref{fig_plot_chi_vs_n0_beta}). With these values, the
thermal contribution from the shocked ambient medium increases far too
much in the inner regions to be hidden by the interstellar
absorption. Therefore the inner zones provide even tighter
constraints: the ambient medium density must in fact be lower than
$0.3 \: \mathrm{cm}^{-3}$.

Our results were obtained from a spectral analysis based on different
components to model the emission from the shock-accelerated particles,
the shocked ejecta and the shocked ambient medium. In particular the
characteristics of the shocked ambient medium were derived from
CR-hydrodynamic models able to match the observed radii ratio between
the blast wave and contact discontinuity as well as the observed
expansion measurements. This is only possible if particle acceleration
is efficient \citep{de05, wah05}.

In that case and for any ambient density, we found from our model that
the range of electronic temperatures allowed in the radial regions
free from strong line emission is between $5 \: \mathrm{keV}$ and $20
\: \mathrm{keV}$ if electron and proton temperatures are equal at the
shock, and between $0.2 \: \mathrm{keV}$ and $2 \: \mathrm{keV}$ if
not (see Fig.~\ref{fig_nei_param_alln0}). This eliminates the
possibility that CR hydro models, which satisfy the observed radii
ratio, could lead to a temperature sufficiently low that the emission
of the shocked ambient medium gets shifted to the extreme UV range.
This does not seem to be a viable explanation for the lack of thermal
X-ray emission from the shocked ambient medium.

We note that our derived upper limit value of $0.3 \:
\mathrm{cm}^{-3}$, which does not depend on the details of the shocked
ejecta emission, is inconsistent (lower by a factor $\sim 3-4$) with
the value found by comparing the observed global thermal X-ray
properties of the shocked ejecta with predictions from various models
of thermonuclear explosions \citep{bab06}.  There are a number of
differences between this published study and the one we present here
(e.g., the density profile of the ejecta, whether or not efficient
shock acceleration is included, whether or not similarity solutions
are used for the hydrodynamics) and both are based on only
one-dimensional hydrodynamics.

If we have underestimated the observed size of the gap between the
blast wave and the contact discontinuity by a factor of two, i.e., if
the ratio of radii were in reality $\sim$1.2 (which is roughly the
value expected in the test particle case for $n=7$, then a higher
ambient medium density would be possible since this would result in a
smaller compression ratio ($r_{\mathrm{tot}} \sim 4$), the ionization
age would increase less rapidly behind the blast wave and the volume
of our regions would slightly decrease (because the distance would be
reduced). But this would be no more than a factor of 2 leading to an
ambient density of $n_{0} \sim 0.4-0.6 \: \mathrm{cm}^{-3}$.  An
overestimate of the gap by a factor 2 (i.e., a ratio of radii of $\sim
1.06$ resulting in $r_{\mathrm{tot}} \sim 10$) would reduce the
estimate on the ambient medium density by the same factor making
$n_{0}$ of order $0.1-0.2 \: \mathrm{cm}^{-3}$. A factor of more than
two error in our estimate of the size of the gap in either direction
is unlikely. A case of lower injection efficiency (e.g.,
$\eta_{\mathrm{inj}} = 2 \times 10^{-4}$) would lead to a higher mean
shock temperature compared to the case with
$\eta_{\mathrm{inj}}=10^{-3}$; this can only lower the density
estimate (see second panel of Fig.
\ref{fig_data_REG_waponei_link_tau_norm_nh0.7}).

 Our use of a power-law initial ejecta density profile is also
  a possible source of uncertainty. By examining density profiles
  generated by thermonuclear SN explosion models
  \citep[e.g.,][]{hok96}, \citet{dwc98} conclude that an exponential
  profile is a reasonably good simple representation for the initial
  ejecta density profile.  When evolving these profiles to the remnant
  stage in a uniform density environment, they find that the power-law
  and exponential profiles produce similar density and temperature
  structures in the shocked ambient medium (while they find large
  differences for the profiles in the shock heated ejecta, which we do
  not study here).  The largest difference between the density
  profiles in the shocked ambient medium \citep[see Fig.~3
  in][]{dwc98} is no more than a factor of two (with the exponential
  case falling below the power-law case).  Both temperature profiles
  are basically flat right behind the blast wave (i.e., where we
  extracted the spectra) and increase more or less rapidly only very
  close to the contact discontinuity (again, a region that we do not
  study).  Although these calculations do not include the effect of
  efficient diffuse shock acceleration, they should be indicating
  roughly the level of difference between the power-law and
  exponential density profiles.  \citet{elc05} have generated radio
  surface brightness profiles for thermonuclear remnants assuming
  power-law and exponential profiles when shock acceleration is
  efficient (see their Fig.~11) from which it is possible to estimate
  how the ratio $R_s/R_c$ differs. Extrapolating their $n=9$ power-law
  results to the $n=7$ power-law ejecta case we used here suggests
  that there is little difference in the ratio of radii compared to
  the exponential case.  In summary, based on admittedly limited
  published results, we estimate that using an exponential ejecta
  density profile could result in an increase in the inferred ambient
  medium density of up to a factor of 2.

\subsection{Origin of the rim morphology\label{subsect-at-rim}}

The objective of the present discussion is to understand why most
of the X-ray synchrotron emission is confined in narrow rims at
the blast wave.

\subsubsection{The two interpretations}

There are currently two alternatives to explain such morphology.  The
first one stipulates that the highest energy electrons cannot travel
(by advection or diffusion) far from their acceleration site -
presumably the blast wave - without suffering from efficient energy
losses due to synchrotron radiation so that their emission is
concentrated to very thin regions \citep{ba03, vil03a}. To produce
enough radiative losses this interpretation requires a rather high
magnetic field within the rims, which presupposes that the field must
have been amplified at the blast wave \citep{vob05, ba06, pam06}. Such
turbulent amplification of the magnetic field in collisionless shock
waves was already suggested on the basis of theoretical investigations
\citep{bel01} and simulations \citep{lub00}. The second possibility
recently suggested by \cite{poy05} is that the observed X-ray rims may
in fact reflect the spatial distribution of the magnetic field rather
than the spatial distribution of the high-energy electrons. This
interpretation assumes also a certain level of amplification of the
magnetic field at the blast wave (or in the precursor) and imposes its
decrease behind, resulting from the relaxation or damping of the
turbulence, on a timescale shorter than the characteristic time for
electrons to lose energy by synchrotron radiation.

The fundamental difference between these two interpretations lies in
the post-shock evolution of the magnetic field (advected or damped) or
equivalently in its ability to modify the energy of the X-ray-emitting
electrons through synchrotron losses and thereby modify their spatial
distribution. Hence, our initial problem on the origin of the rim
morphology is nothing more than a problem related to the evolution of
the magnetic field behind the blast wave. In the following, we will
attempt to determine the magnetic field characteristics/properties
(i.e., its intensity at the shock and behind) and the parameters of
the acceleration (i.e., the injection efficiency, the maximum energy
of the shock-accelerated electrons, and the density ratio between the
relativistic electrons and protons), by comparing the associated
modelled synchrotron emission properties (i.e., brightness and photon
index) with those derived from the observations. There are a large
number of observational constraints which allow us to strongly
constrain the previous parameters: the ratio of radii between the
blast wave and contact discontinuity, the width of the X-ray rims, the
X-ray spectral variations behind the blast wave (\S
\ref{subsect-PLindex}), the radio and X-ray brightness (\S
\ref{subsect-radio-xray}), and the upper limit on the ambient medium
density (\S \ref{subsect-sam}).

\subsubsection{CR-hydro model and particle spectra\label{subsect-CR-hydro+spectra}}

To compute the properties of the synchrotron emission, we take
advantage of the CR-modified hydrodynamic model of SNR evolution
that we used previously to estimate the density in the ambient medium
(see \S \ref{subsect-sam}).

For a given CR injection efficiency, $\eta_{\mathrm{inj}}$, and
ambient density, $n_0$, the CR-hydro model provides the CR proton
spectrum at the blast wave at any time. It is a piece-wise
  power-law model with an exponential cutoff at high energies:
\begin{equation} \label{fp}
f_{\mathrm{p}}(E) = a \:  E^{-\Gamma(E)} \:
 \exp \left( -
 E / E_{\mathrm{p,max}} \right),
\end{equation}
where $a$ is the normalization, $\Gamma$ is the power-law index which
depends on the energy E, and $E_{\mathrm{p,max}}$ is the maximum
energy reached by the protons. Typically three distinct energy regimes
with different $\Gamma$ values are assumed \citep{bee99}.  The
normalization, $a$, is proportional to $\eta_{\mathrm{inj}}$ and
$n_0$. The CR electron spectrum is determined by assuming a certain
electron-to-proton density ratio at relativistic energies,
$K_{\mathrm{ep}}$, which is defined as the ratio between the electron
and proton distributions at a regime in energy where the protons are
already relativistic but the electrons have not yet cooled radiatively
\citep[e.g.,][]{elb00}. In the appropriate energy range, the
  CR electron spectrum is then:
\begin{equation} \label{fe}
f_{\mathrm{e}}(E) = a \: K_{\mathrm{ep}} \: E^{-\Gamma(E)} \:
 \exp \left( - 
 E / E_{\mathrm{e,max}} \right),
\end{equation}
where $E_{\mathrm{e,max}}$ is the maximum energy reached by the
electrons.  $K_{\mathrm{ep}}$ is left as a free parameter \citep[e.g.,
][]{elb00, vob02} and will be adjusted using the X-ray data (see \S
\ref{subsect-methodology}). We will obtain typically $K_{\mathrm{ep}}
\sim 10^{-3}$ (see \S \ref{subsect-synchproj-profile}).

The maximum energies $E_{\mathrm{p,max}}$ and $E_{\mathrm{e,max}}$
contain information on the limits of the acceleration. They are set by
matching either the acceleration time to the shock age or to the
characteristic time for synchrotron losses, or by matching the
upstream diffusive length to some fraction $\xi_{s}$ of the shock
radius (i.e., escape limitation), whichever gives the lowest value. We
took $\xi_{s} = 0.05$. When the maximum energy of the electrons is
limited by radiative losses, we have:
\begin{equation}\label{Ee,max}
E_{\mathrm{e,max}} = \frac{3 \: m_{\mathrm{e}}^2 \: c^3}{2 \:
e^{3/2}} \: \sqrt{ \frac{(r-1)/r}{r + 1/r_{\mathrm{B}}} } \:
k_{0}^{-1/2} \: B_{2}^{-1/2} \: V_{s},
\end{equation}
where $r$ and $r_{\mathrm{B}}$ are the overall density and magnetic
field compression ratios, $B_2$ the immediate post-shock value of the
magnetic field, $V_s$ the shock speed, and $k_{0} \equiv
D(E_{\mathrm{e,max}}) / D_{\mathcal{B}}(E_{\mathrm{e,max}})$ the ratio
between the diffusion coefficient, $D$, and its Bohm value,
$D_{\mathcal{B}}$, both at $E_{\mathrm{e,max}}$
\citep[see][]{pam06}. With typically $r \simeq 6$ and $r_{\mathrm{B}}
\simeq 5$ (see \S \ref{subsect-MF-profiles}), we obtain:
\begin{equation}\label{Ee,max-2}
E_{\mathrm{e,max}} \simeq 7.3 \: k_{0}^{-1/2} \: B_{100}^{-1/2} \:
V_{s,3} \: \mathrm{TeV},
\end{equation}
where $B_{100}$ is $B_2$ in units of $100 \: {\mu}\mathrm{G}$, and
$V_{s,3}$ is $V_{s}$ in units of $1000 \: \mathrm{km/s}$.  In
Eq. (\ref{fe}), $E_{\mathrm{e,max}}$ is left as a free parameter (via
$k_0$) and will be adjusted using the X-ray data (see \S
\ref{subsect-methodology}).  We will obtain typically $k_0 \sim 10$
making $E_{\mathrm{e,max}} \sim 10 \: \mathrm{TeV}$.  Finally, for
simplicity, we assumed the Bohm value and regime for the diffusion
coefficient of relativistic protons.  Because the protons affect the
modelling only via their total energy density, our results are not
very sensitive to $E_{\mathrm{p,max}}$.

Once the particle distributions associated with their respective
fluid elements are produced, they evolve downstream experiencing
adiabatic and eventually synchrotron losses (as described in
Appendix \ref{app-slope-losses}). Then, using the electron
distributions, the synchrotron emission within the remnant can be
calculated, provided that the magnetic field structure is known
within the remnant.

\subsubsection{Magnetic field profiles\label{subsect-MF-profiles}}

We present four configurations of the downstream magnetic field as a
function of radius (normalized to the contact discontinuity) as
illustrated in Figure \ref{fig_bmag_profile}: two where the magnetic
field is damped (left panels), and two where the magnetic field is
simply advected behind the shock, i.e., it is passively carried by
the plasma flow (right panels). Appendix \ref{app-MF-profiles}
explains how we evolve the magnetic field associated with each fluid
element and then how the magnetic field profiles were
obtained. Anticipating the results we obtain below, for each magnetic
field behavior (damped or advected), we show a case with an injection
efficiency, $\eta_{\mathrm{inj}}$, of $10^{-3}$ and pre-shock magnetic
field, $B_0$, of $\sim 45 \: \mu{\mathrm{G}}$, and one with a lower
injection efficiency of $1.4 \times 10^{-4}$ and lower but still high
pre-shock magnetic field of $\sim 25 \: \mu{\mathrm{G}}$.  Unless
explicitly stated, these two cases are always given for an ambient
density, $n_0$, of $0.2 \: \mathrm{cm}^{-3}$ and a kinetic energy of
the explosion, $E_{\mathrm{SN}}$, of $10^{51} \: \mathrm{ergs}$.  The
case with $\eta_{\mathrm{inj}} = 10^{-3}$ produces very efficient
diffusive shock acceleration, and the case with $\eta_{\mathrm{inj}} =
1.4 \times 10^{-4}$ yields a less efficient acceleration but still
with some fraction of the energy flux crossing the shock going into
relativistic particles and where the back reaction of
shock-accelerated protons on the hydrodynamics is still important.

The values obtained for the unshocked magnetic field, $B_{0}$, are
clearly several times higher than the typical interstellar magnetic
field of a few $\mu{\mathrm{G}}$. We implicitly assume that $B_{0}$
has been already significantly amplified by some instabilities
provided, for example, by the streaming of accelerated CRs in the
precursor \citep[see][]{bel01, pel06, mal06}. Assuming the magnetic
turbulence to be isotropic ahead of the shock, the magnetic field
downstream is then larger than upstream by a factor $r_{\mathrm{B}} =
\sqrt{(1+ 2 \: r_{\mathrm{tot}}^2)/3}$. Since the ratio of radii
between the blast wave and contact discontinuity in the W rim
constrains the overall compression ratio, $r_{\mathrm{tot}}$, to a
value of 6 (see \S \ref{subsect-cr-hydro-model}), we have
$r_{\mathrm{B}} \simeq 5$. This leads to an immediate post-shock
magnetic field, $B_2$, equal to $\sim 215 \: \mu{\mathrm{G}}$ when
$\eta_{\mathrm{inj}} = 10^{-3}$ and $\sim 130 \: \mu{\mathrm{G}}$ when
$\eta_{\mathrm{inj}} = 1.4 \times 10^{-4}$. These two sets of
$(\eta_{\mathrm{inj}},B_2)$ will allow us to describe fairly well the
X-ray data, i.e., the intensity of the X-ray rims (assuming a
reasonable $K_{\mathrm{ep}}$ ratio), their width and the spatial
variations of the X-ray photon index.

Figure \ref{fig_bmag_profile} shows that, when the magnetic field
is damped behind the shock (left panels), the final profile
obtained at an age of $430$ years is roughly exponentially
decreasing, producing a magnetic filament at the blast wave. The
higher the injection efficiency and immediate post-shock magnetic field, 
the sharper the filament (see Appendix \ref{app-MF-damped}). 
When the magnetic field is advected
behind the shock (right panels), the final magnetic field profile
is also decreasing but far less so than in the damped case. These
differences in the magnetic field profile/evolution will lead to
different characteristics of the synchrotron emission.

\subsubsection{Synchrotron spectra}

Figure \ref{fig_syn_spectra} shows the synchrotron spectrum generated
at the blast wave and those produced by several fluid elements
corresponding to different observed zones (whose flow times are given
in Table \ref{tab-CR+hydro-simu}) at the remnant's age of 430 years.
Note that the calculated synchrotron emissivity, $\epsilon_{\nu}$, was
averaged over viewing angles. Our study does not include the effect of
magnetic field orientation.  The different panels correspond to our
different assumptions about the magnetic field evolution (damped in
the left panels, advected in the right panels) and energy losses (only
adiabatic expansion losses\footnote{In that case, the slope of the
  synchrotron spectrum reflects directly the slope of the electron
  spectrum when it was produced at the shock (see appendix
  \ref{app-slope-losses}).} in the top panels, adiabatic expansion
plus radiative losses in the bottom panels). In those examples, the
injection efficiency and immediate post-shock magnetic field were
fixed to $1.4 \times 10^{-4}$ and $130 \: {\mu}\mathrm{G}$, and
$K_{\mathrm{ep}}$ to $10^{-3}$ and $k_0$ to $10$ (left panels) or $7$
(right panels).

We illustrate the crucial role of the magnetic field (both its
strength and evolution) on the production of the synchrotron
emission. When the magnetic field is damped (left panels), the
spectral variations of the different photon spectra come from
variations in the strength of the magnetic field which shift the
emission in frequency/energy as illustrated by the case with only
adiabatic losses included (top-left panel). In the case of
efficient particle acceleration considered here
($\eta_{\mathrm{inj}} = 1.4 \times 10^{-4}$ and $B_2 = 130 \:
{\mu}\mathrm{G}$), the magnetic field strength diminishes rapidly
as we move in from the blast wave to the interior. However,
because the magnetic field is very large at the blast
wave, its cumulative effect over time leads to substantial
synchrotron losses and then clear changes in the synchrotron
spectrum slope (bottom-left panel).

On the other hand, when the magnetic field is carried by the
plasma flow (right panel of Fig.~\ref{fig_syn_spectra}),
differences between the case with only adiabatic losses and the
one with adiabatic plus synchrotron losses are even more
important. Because the magnetic field strength stays approximately
constant behind the blast wave, the different spectra are almost
unshifted in energy and therefore very similar, but in turn this
produces very strong synchrotron losses and therefore significant
spectral variations. These spectral variations are much more
pronounced than those obtained for a decreasing magnetic field
(compare the bottom panels).

\subsubsection{Methodology\label{subsect-methodology}}

Now that the details of how we calculate the magnetic field profiles
and synchrotron spectra are established, here we outline our
methodology for relating relevant model parameters to observational
constraints. The starting point is the ratio of radii, $R_s/R_c$,
between the contact discontinuity and the blast wave, equal to 1.113
here in the W rim (see \ref{subsect-BW-CD}). This provides a relation
between the injection efficiency, $\eta_{\mathrm{inj}}$, the
post-shock magnetic field, $B_{2}$, and the ambient density,
$n_0$. The kinetic energy of the explosion, $E_{\mathrm{SN}}$, should
be considered as a free parameter that also influences
$R_s/R_c$. However, we freeze $E_{\mathrm{SN}}$ at $10^{51} \:
\mathrm{ergs}$, for simplicity.  Figure \ref{fig_RsoRc_etainj} shows
for instance how the ratio of radii, $R_s/R_c$, varies as a function
of $B_{2}$ for several values of $\eta_{\mathrm{inj}}$ at fixed $n_0$
(left panel) or how it varies as a function of $n_0$ for several
values of $\eta_{\mathrm{inj}}$ at fixed $B_2$ (right panel).

We select a value of $n_0=0.2 \: \mathrm{cm}^{-3}$ which is fully
consistent with the lack of shocked thermal ambient medium and derive
the injection value consistent with the ratio of radii for a given
value of magnetic field (left panel of Fig~\ref{fig_RsoRc_etainj}).
Then we calculate the projected profiles of the X-ray synchrotron
brightness and spectral index, and verify that these model profiles
are consistent with the observations. At this point, we are not making
a comparison to the surface brightness data in flux units, but rather
only to the normalized profile. The process is iterated using
different values of $B_{2}$ in order to bracket the range of X-ray
profile widths. We obtain reasonable fits for each magnetic field
configuration (damped or advected) and this effectively results in a
constraint on $B_{2}$.  At the same time, we can calculate the
normalized radio profile as well as the ratio of X-ray to radio
fluxes. The most important parameter which governs that ratio is the
maximum energy reached by the electrons, $E_{\mathrm{e,max}}$. We set
$E_{\mathrm{e,max}}$ (via $k_0$) in order to get the right average
X-ray spectral slope. This is done for each magnetic field
configuration and, as we will see below, the modelled radio profiles
are quite different. Finally we determine the electron-to-proton
density ratio, $K_{\mathrm{ep}}$, by scaling the modelled X-ray
profile to the peak values of the \textit{Chandra} data at the
rim. Reasonable values of order $10^{-3}$ for $K_{\mathrm{ep}}$ will
be obtained.

\subsubsection{Radial profiles of the synchrotron emission\label{subsect-synchproj-profile}}

\begin{figure}[t]
\centering
\includegraphics[width=8cm]{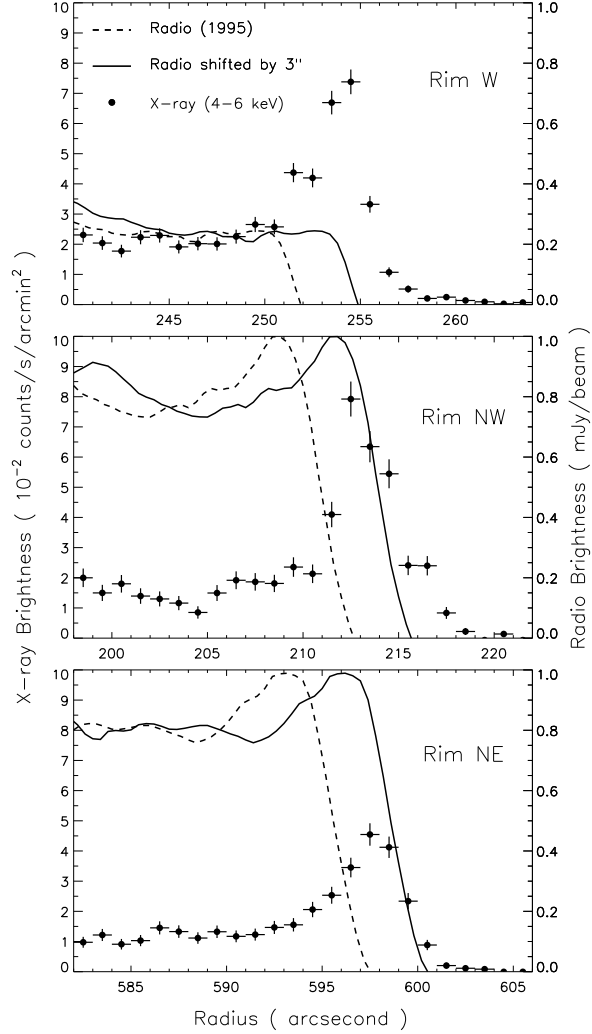}
\caption{Comparison between radio and X-ray radial profiles of the
  rims W (\textit{top panel}), NW (\textit{middle panel}) and NE
  (\textit{bottom panel}). The X-ray profiles were made from count
  images of the 4-6 keV continuum emission ({\Large $\bullet$} data
  points, scale on the left). The radio profiles (dashed line, scale
  on the right) were obtained by using the 1994-1995 \textit{VLA}
  high-resolution ($\sim 1\arcsec$) radio data
  \citep[see][]{rem97}. Errors on the amplitude of the radio profiles
  are about $10 \%$. These profiles were shifted by $3\arcsec$ (solid
  line) to be comparable with the 2003 \textit{Chandra} X-ray
  data.} \label{fig_radio-xray_profiles}
\end{figure}

\begin{figure*}[t]
\centering
\includegraphics[width=14cm]{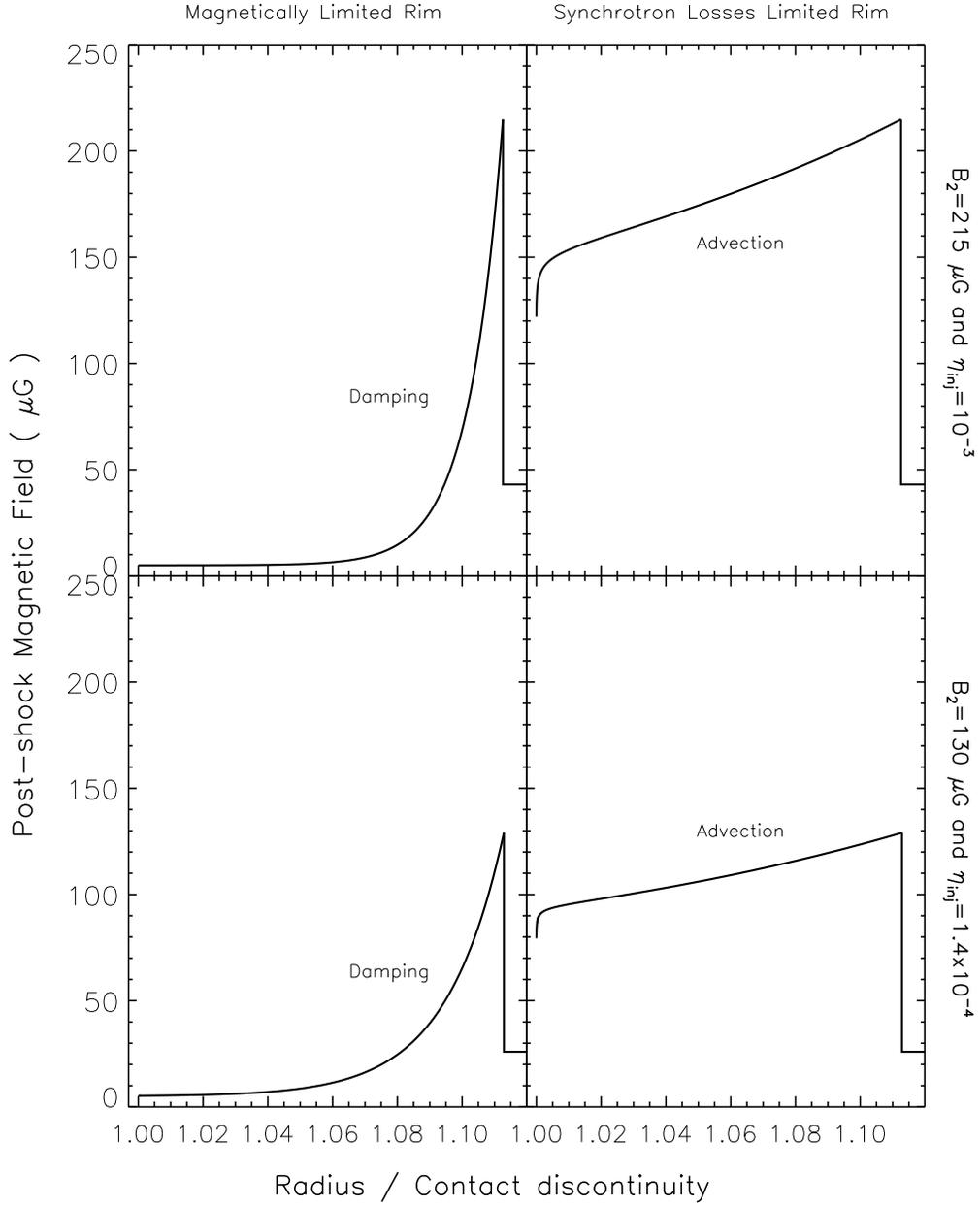}
\caption{Four post-shock magnetic field configurations that we
consider to interpret the observed spatial and spectral variations
of the synchrotron emission: two resulting from the damping of the
turbulence (\textit{left panels}) and two from the advection of
the shocked plasma (\textit{right panels}). We show the profiles
for two couples of injection efficiency and immediate post-shock
magnetic field: $\eta_{\mathrm{inj}} = 10^{-3}$ and $B_2 = 215 \:
{\mu}\mathrm{G}$ (\textit{top panels}) and $\eta_{\mathrm{inj}} =
1.4 \times 10^{-4}$ and $B_2 = 130 \: {\mu}\mathrm{G}$
(\textit{bottom panels}).} \label{fig_bmag_profile}
\end{figure*}

\begin{figure*}[t]
\centering
\includegraphics[bb=0 40 566 850,clip,width=13cm]{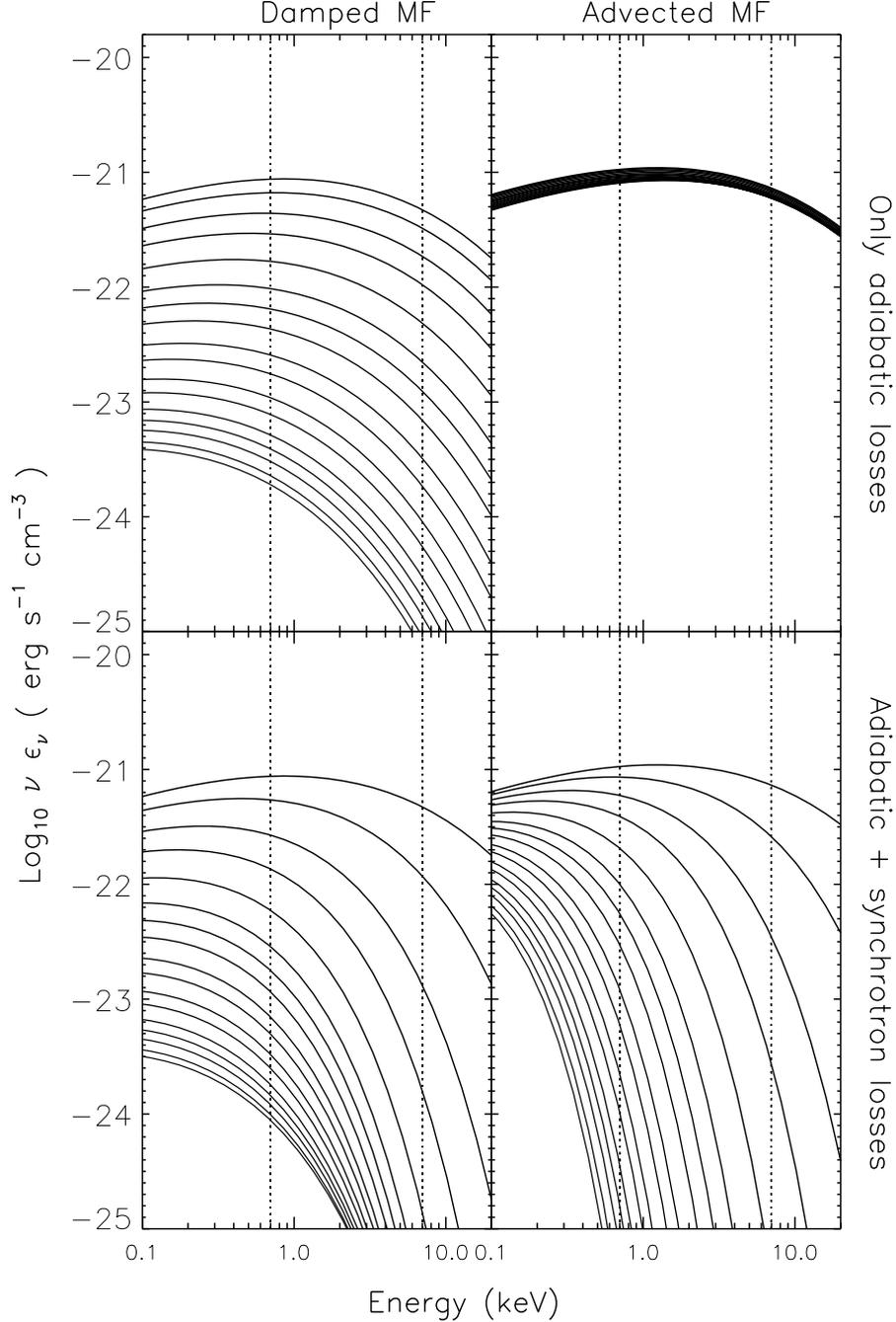}
\caption{Synchrotron spectra produced at various places within the
remnant corresponding to the flow time $\Delta t = 0$ and those
indicated in Table \ref{tab-nei-hydro+ej-xspec-fit}. The four
panels correspond to different assumptions made on the magnetic
field (MF) evolution/profile and nature of the energy losses. The
injection efficiency, $\eta_{\mathrm{inj}}$, was fixed to $1.4
\times 10^{-4}$ and the immediate post-shock magnetic field,
$B_2$, to $130 \: {\mu}\mathrm{G}$. The dashed lines indicate the
0.7-7 keV energy band.} \label{fig_syn_spectra}
\end{figure*}

\begin{figure*}[t]
\centering
\begin{tabular}{cc}
\includegraphics[width=8cm]{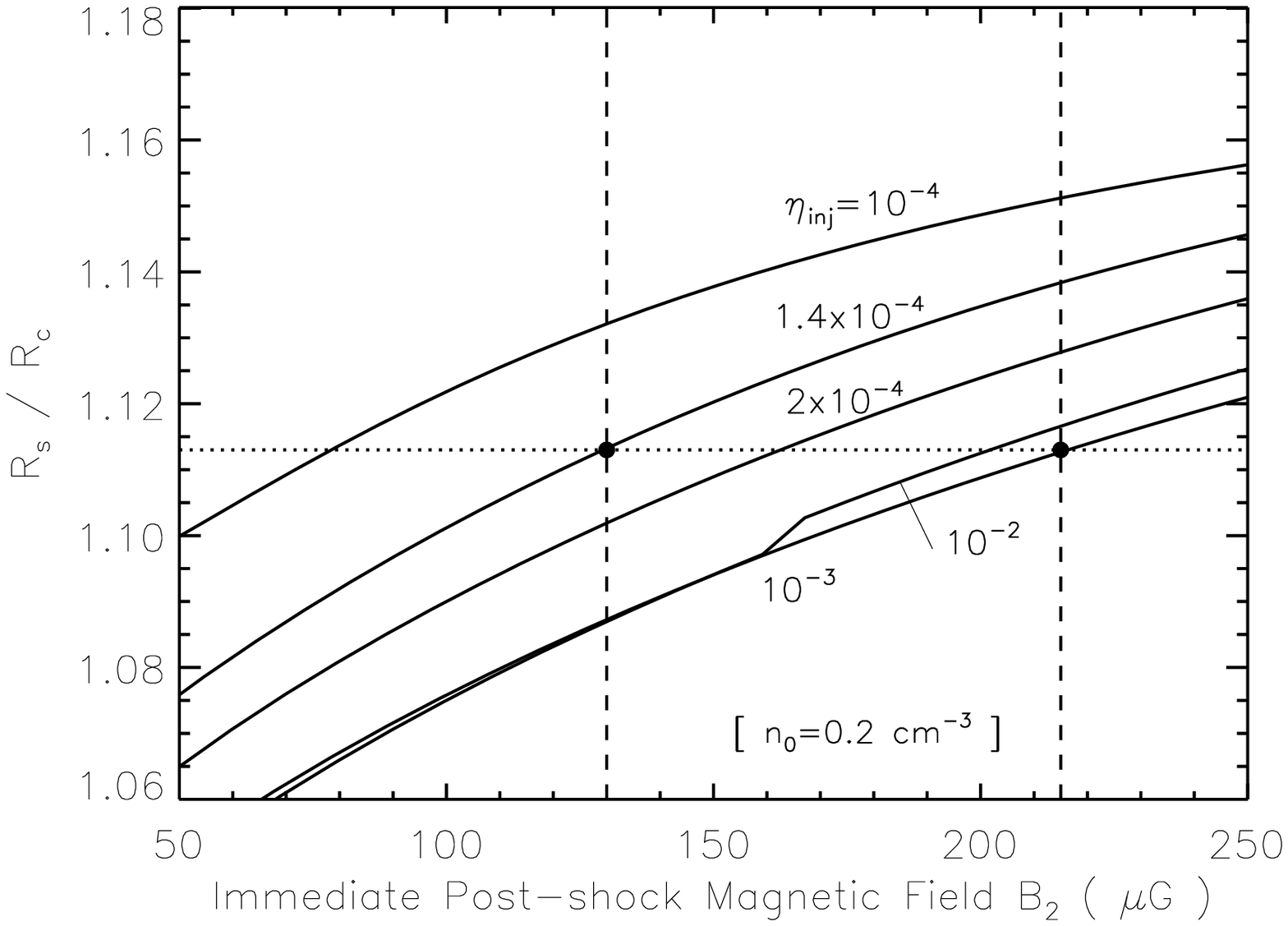} &
\includegraphics[width=8cm]{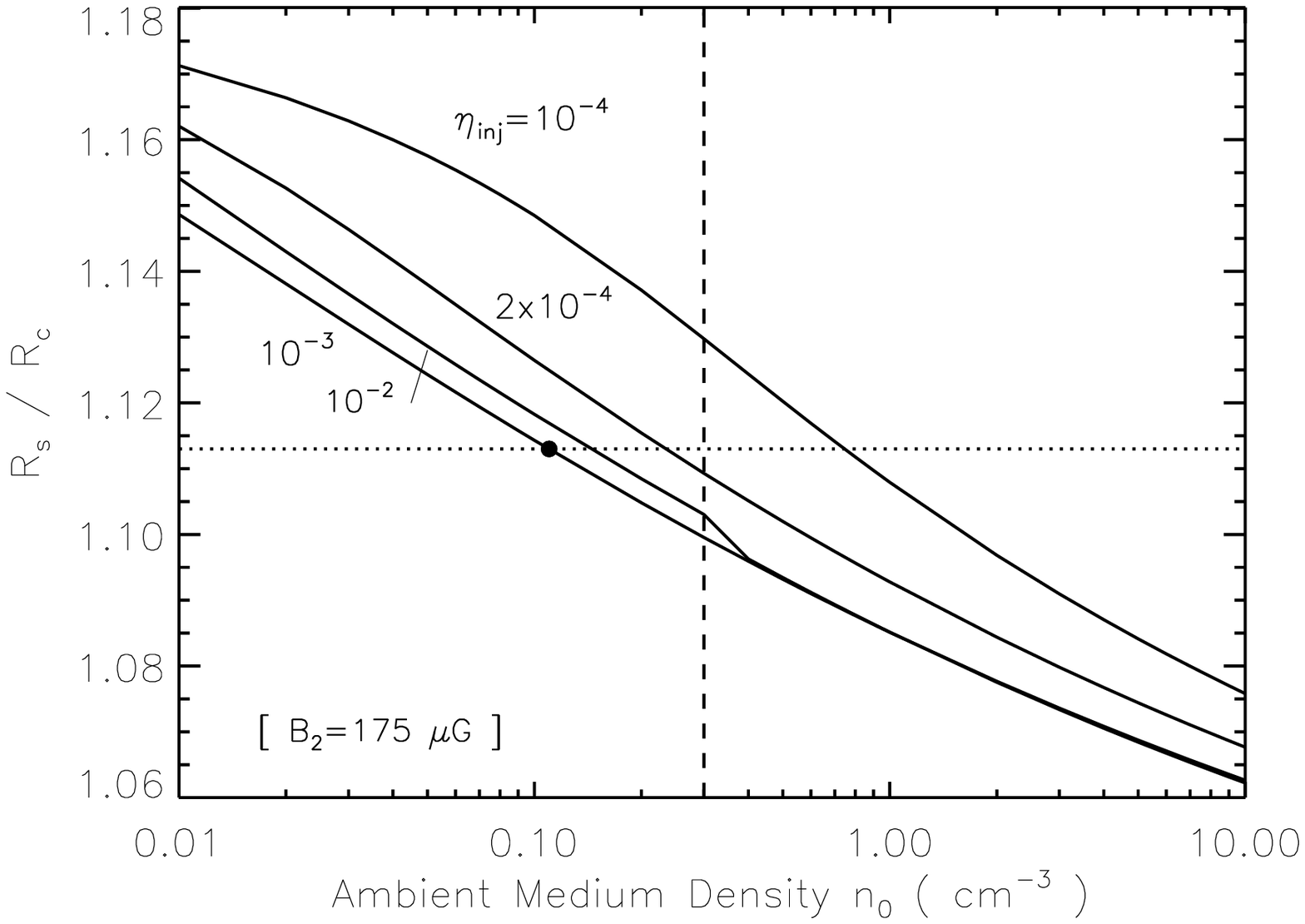}
\end{tabular}
\caption{\textit{Left panel:} predicted ratio of radii, $R_s/R_c$,
between the blast wave and the contact discontinuity as a function
of the immediate post-shock magnetic field, $B_2$, for different
values of the injection efficiency, $\eta_{\mathrm{inj}}$, with a
pre-shock ambient density, $n_{0}$, fixed to $0.2 \:
\mathrm{cm}^{-3}$. The horizontal dotted line indicates the ratio
of radii measured in the W rim of Tycho. On this line, the overall
compression ratio is equal to 6. The two {\Large $\bullet$} are
the values that we chose to illustrate in Figures
\ref{fig_bmag_profile}, \ref{fig_synchproj_profile} and
\ref{fig_synchproj_slope}. Above $B_2 \simeq 160 \:
{\mu}\mathrm{G}$, the curve with $\eta_{\mathrm{inj}} = 10^{-2}$
is above that with $\eta_{\mathrm{inj}} = 10^{-3}$. \textit{Right
panel:} same but as a function of $n_{0}$ with $B_2$ fixed to $175
\: \mu{\mathrm{G}}$. The dashed line indicates the upper limit of
$0.3 \: \mathrm{cm}^{-3}$ found from the lack of thermal X-ray
emission from the shocked ambient medium. The {\Large $\bullet$}
indicates the lower limit on the ambient density ($\sim 0.1 \:
\mathrm{cm}^{-3}$) allowed by the CR-hydro model. Below $n_0
\simeq 0.4 \: \mathrm{cm}^{-3}$, the curve with
$\eta_{\mathrm{inj}} = 10^{-2}$ is above that with
$\eta_{\mathrm{inj}} = 10^{-3}$.} \label{fig_RsoRc_etainj}
\end{figure*}

\begin{figure*}[t]
\centering
\includegraphics[width=14cm]{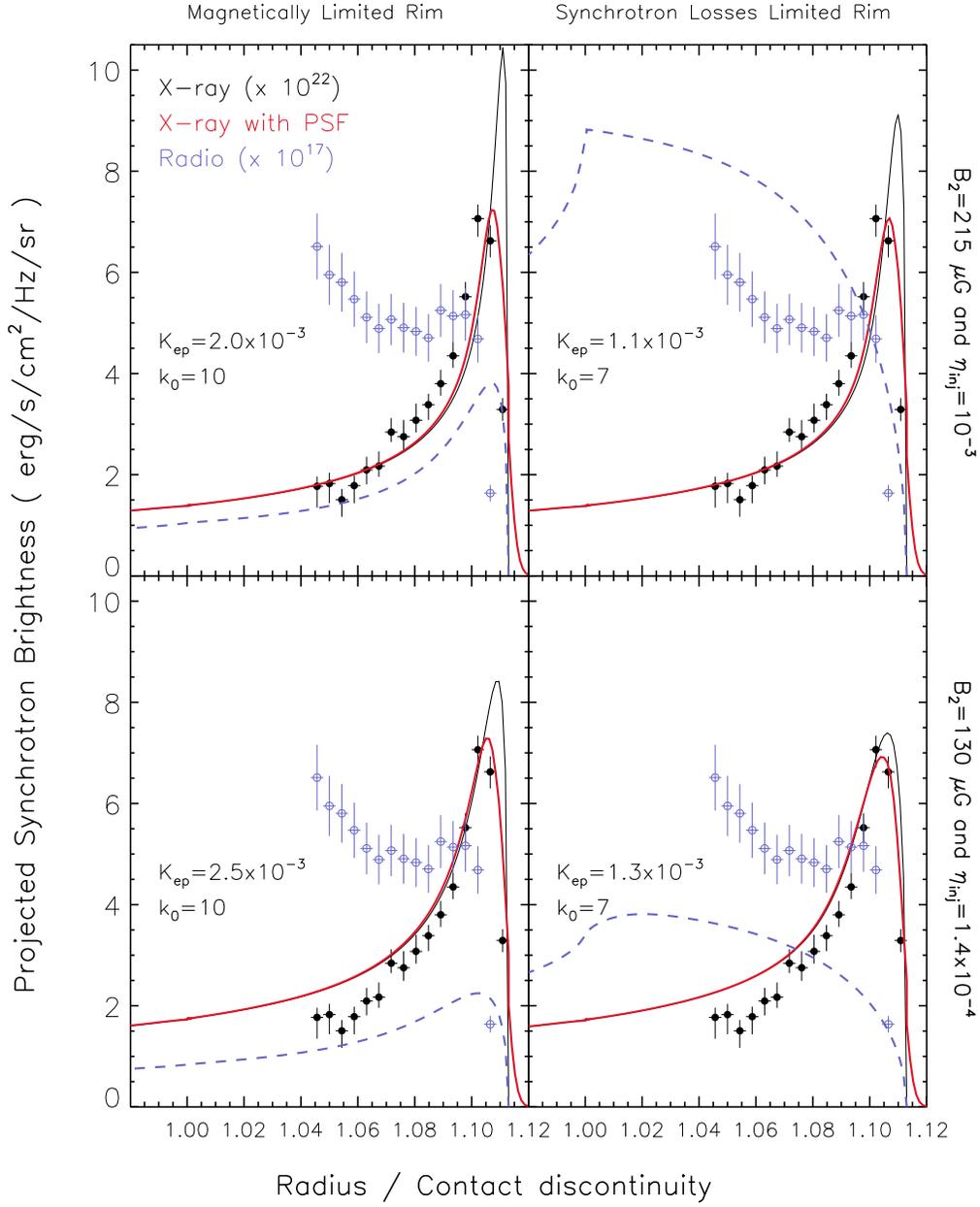}
\caption{Line-of sight projections of the modelled radio (1.4 GHz,
  blue dotted lines) and X-ray (1 keV, black solid lines) synchrotron
  brightness obtained for the post-shock magnetic field configurations
  shown in Fig. \ref{fig_bmag_profile}. The red solid lines show the
  X-ray profiles convolved with a model of the \textit{Chandra}
  PSF. The radio (marked with {\Large $\circ$}, in blue) and X-ray
  (marked with {\Large $\bullet$}) data points of the W rim are shown
  for comparison. The electron-to-proton density ratio at relativistic
  energies, $K_{\mathrm{ep}}$, was adjusted so that the model (red
  lines) matches the intensity of the X-ray rim.  The radio and X-ray
  profiles were multiplied by $10^{17}$ and $10^{22}$, respectively,
  to make them comparable on the plot.}
\label{fig_synchproj_profile}
\end{figure*}

In Figure \ref{fig_synchproj_profile}, we plot the expected
line-of-sight projections of the synchrotron brightness\footnote{That
  is $( 4 \: \pi )^{-1} \: \int \epsilon_{\nu}(r) \: dl$ performed
  along the line-of-sight and where $\epsilon_{\nu}$ is the
  synchrotron emissivity per unit volume at the frequency $\nu$ (in
  erg/s/cm$^{-3}$/Hz) which appears for instance in Figure
  \ref{fig_syn_spectra}.} (with both adiabatic and synchrotron losses
included) in one radio (blue dotted lines) and one X-ray band (black
solid lines) corresponding to the four magnetic field configurations
shown in Figure \ref{fig_bmag_profile}. In addition, we show the X-ray
profiles convolved with a gaussian that matches the \textit{Chandra}
PSF at 1 keV (red solid lines) which allows us a direct comparison
with the X-ray data (points marked with {\Large $\bullet$}). The X-ray
data points correspond to the power-law normalization obtained by
fitting the \textit{Chandra} spectra in the W rim with a power-law
plus an ejecta template and a fixed absorption of $0.7 \times 10^{22}
\: \mathrm{cm}^{-2}$ (see \S \ref{subsect-pl+ej}), and divided by the
solid angle of each radial bin. The predicted radio profiles can be
compared with the radio data points (marked with {\Large $\circ$} in
blue) which differs from those of Figure \ref{fig_radio-xray_profiles}
(top panel) as they account for the beam size. For projection we
assume the blast wave curvature in the out-of-sky-plane direction to
be equal to the sky-plane curvature.

When the magnetic field is damped behind the blast wave (left panels
of Fig.~\ref{fig_synchproj_profile}), the modelled radio (dotted
lines) and X-ray (red solid lines) morphologies are very similar. Both
radio and X-ray profiles are peaked at the blast wave, with the radio
profile wider and with a smaller brightness contrast between the rim
and far behind the rim. The limb-brightening of the modelled radio
profiles comes closer to matching what we observe in the radio rims in
terms of morphology (see the NW and NE rims in
Fig.~\ref{fig_radio-xray_profiles}), but clearly not in terms of
intensity. When either $\eta_{\mathrm{inj}} = 10^{-3}$ and $B_2 = 215
\: {\mu}\mathrm{G}$ (top-left panel) or $\eta_{\mathrm{inj}} = 1.4
\times 10^{-4}$ and $B_2 = 130 \: {\mu}\mathrm{G}$ (bottom-left
panel), the predicted radio brightness is far too low compared to the
radio data. In the magnetic damping case, once $E_{\mathrm{e,max}}$
(or equivalently $k_0$) has been fixed in order to be consistent with
the observed average X-ray slope, it is difficult to increase the
radio intensity while keeping the same level of X-ray intensity
because the ratio of X-ray to radio fluxes (independent of
$K_{\mathrm{ep}}$) does not depend strongly on the injection
efficiency. However, a slight variation in $k_0$ can increase the
radio emission while keeping the same level of X-ray emission
(provided that we adjust $K_{\mathrm{ep}}$) and without changing the
average X-ray slope too much.  For instance, when $\eta_{\mathrm{inj}}
= 10^{-3}$ and $B_2 = 215 \: {\mu}\mathrm{G}$ (top-left panel),
changing $k_0$ to 14 (instead of 7) and adjusting $K_{\mathrm{ep}}$ to
$2.5 \times 10^{-3}$ (instead of $2.0 \times 10^{-3}$) increases the
radio emission by a factor 1.25 (which corresponds to the ratio of
$K_{\mathrm{ep}}$) without modifying the X-ray intensity and only
increases the averaged X-ray photon index by 0.1 (index of 2.9 instead
of 2.8).  Note however that in the case $\eta_{\mathrm{inj}} = 1.4
\times 10^{-4}$ and $B_2 = 130 \: {\mu}\mathrm{G}$ (bottom-left
panel), it is not possible to make the absolute radio and X-ray fluxes
at the rim and the average X-ray photon index all consistent with the
observations just by varying $k_0$.  It requires us to change the
shape of the electron cutoff.  We investigate this possibility below
(see \S \ref{subsect-dependencies}).

When the magnetic field is advected behind the blast wave (right
panels of Fig.~\ref{fig_synchproj_profile}), the modelled radio
(dotted lines) and X-ray (red solid lines) morphologies are very
different. Contrary to the X-ray profile which is strongly peaked just
behind the blast wave, the radio profile rises slowly to a maximum
near the contact discontinuity (assuming no additional contribution
from the ejecta). The case $\eta_{\mathrm{inj}} = 1.4 \times 10^{-4}$
and $B_2 = 130 \: {\mu}\mathrm{G}$ with $K_{\mathrm{ep}}=1.3 \times
10^{-3}$ (bottom-right panel) predicts a lower radio intensity
compared to the radio data while the case $\eta_{\mathrm{inj}} =
10^{-3}$ and $B_2 = 215 \: {\mu}\mathrm{G}$ with $K_{\mathrm{ep}}=1.1
\times 10^{-3}$ (top-right panel) predicts a larger radio
intensity. This suggests, however, that there is an intermediate value
for the injection efficiency that can account for the observed level
of radio emission.  We found that $\eta_{\mathrm{inj}} = 3 \times
10^{-4}$ with $B_{2} \simeq 175 \: {\mu}\mathrm{G}$ and
$K_{\mathrm{ep}} = 1.2 \times 10^{-3}$ would provide such a good fit
(not shown).  Nevertheless, even with these best-fit parameters, we
note that the observed rapid rise of the radio emission at the blast
blast can not be reproduced accurately by the models.  Finally, for
completeness, we note that other studies based only on the integrated
synchrotron spectrum of Tycho from the radio to the X-ray suggest a
$K_{\mathrm{ep}}$ value of $4 \times 10^{-3}$, roughly consistent with
ours, although it was obtained with a different set of parameters
\citep{vob02}.

\subsubsection{Radial variations of the X-ray photon index\label{subsect-proj-ph-profile}}

\begin{figure*}[t]
\centering
\includegraphics[width=14cm]{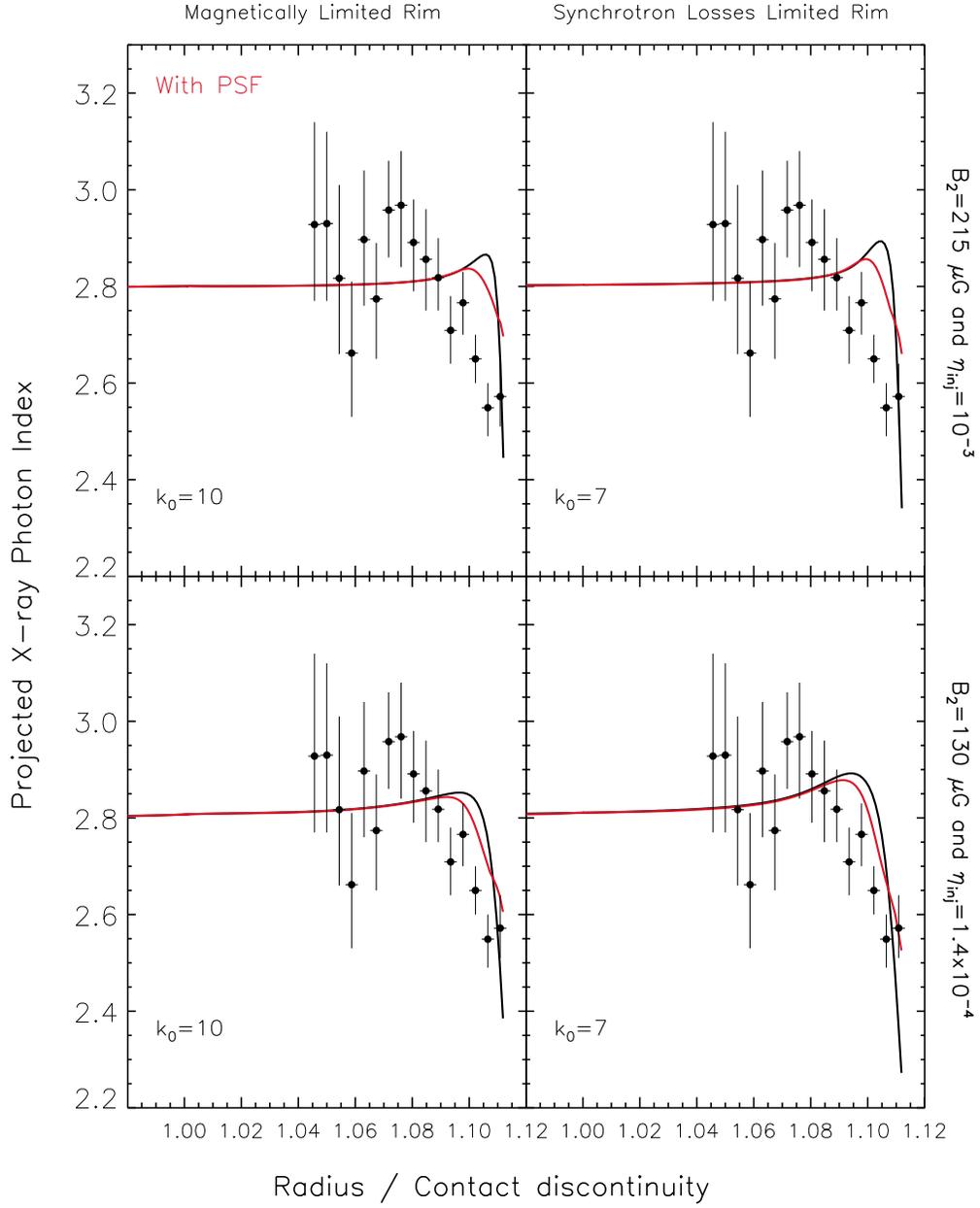}
\caption{X-ray photon index profiles associated with the radial
distributions shown in Fig. \ref{fig_synchproj_profile}. We show the
profiles with the \textit{Chandra} PSF effects taken into account (red
lines), and, for comparison, the \textit{Chandra} data points of the W
rim (marked with {\Large $\bullet$}). The curves are shown with $k_{0}
= 10$ (left panels) and $k_{0} = 7$ (right panels) where $k_{0}$
appears in Eq. (\ref{Ee,max}) (see \S
\ref{subsect-CR-hydro+spectra}).} \label{fig_synchproj_slope}
\end{figure*}

In Figure \ref{fig_synchproj_slope}, we plot the predicted X-ray
photon index as a function of radius corresponding to the four
magnetic field configurations shown in Figure
\ref{fig_bmag_profile}. To obtain these profiles, we had to
construct the projected synchrotron spectrum at each radial
position using a set of projected profiles of the synchrotron
emission computed at several energies within the X-ray band
($0.7-7$ keV), with both adiabatic and radiative losses included.
We show in addition the photon index profiles obtained from a set
of projected brightness profiles at different energies convolved
with a gaussian that matches the \textit{Chandra} PSF at the
appropriate energy (red solid lines). For comparison, we have
plotted as data points the X-ray photon index obtained when
fitting the rim W spectra with a power-law plus an ejecta template
and a fixed absorption of $0.7 \times 10^{22} \: \mathrm{cm}^{-2}$
(see \S \ref{subsect-pl+ej}). Each curve in Figure
\ref{fig_synchproj_slope} tells us how the projected synchrotron
morphology changes with energy/frequency.

Figure \ref{fig_synchproj_slope} shows that projected X-ray photon
index corresponding to the four combinations of magnetic field
evolution can fairly well describe the average X-ray photon index
measured in the W rim. A good match to the average X-ray photon
index profile is obtained with $k_0 =10$ in the magnetic damping
case and $k_0=7$ in the synchrotron losses case, where $k_0$ is
defined in Eq.~(\ref{Ee,max}) (\S \ref{subsect-CR-hydro+spectra}).
We note that the four cases do predict different overall ranges in
the photon index radial profile (red solid lines). The magnetic
damping case (left panels) shows a range of $\sim 0.15$ when
$\eta_{\mathrm{inj}} = 10^{-3}$ and $B_2 = 215 \:
{\mu}\mathrm{G}$, and $\sim 0.25$ when $\eta_{\mathrm{inj}} = 1.4
\times 10^{-4}$ and $B_2 = 130 \: {\mu}\mathrm{G}$. The range is
somewhat larger for the synchrotron losses limited case (right
panels), where the magnetic field is advected behind the blast
wave: $\sim 0.20$ when $\eta_{\mathrm{inj}} = 10^{-3}$ and $B_2 =
215 \: {\mu}\mathrm{G}$, and $\sim 0.35$ when $\eta_{\mathrm{inj}}
= 1.4 \times 10^{-4}$ and $B_2 = 130 \: {\mu}\mathrm{G}$. In fact
these ranges are roughly consistent with the range observed ($\sim
0.3-0.4$), at least for the case $\eta_{\mathrm{inj}} = 1.4 \times
10^{-4}$ and and $B_2 = 130 \: {\mu}\mathrm{G}$. We note however
that the spatial variations of the modelled photon index occur
over a very narrow region close to the shock, although the
gradient of photon index is slightly broadened when the PSF is
included (red solid lines).

In fact, none of the curves is able to reproduce precisely the scale
length of the observed spectral variations without introducing an
inconsistency between the morphology of the projected synchrotron
emission and the observed one. It can be demonstrated (in the thin
spherical shell approximation) that most of the predicted spectral variation
occurs over a distance corresponding to something a little less than
the distance between the blast wave and the location where the
projected brightness falls to half its maximum value (see Appendix
\ref{app-ph-index-profile}). This scale length is shorter than that
over which the observed photon indices appear to increase at least for
rim W.

\subsubsection{Dependencies on model assumptions\label{subsect-dependencies}}

In the two previous sections \S \ref{subsect-synchproj-profile} and \S
\ref{subsect-proj-ph-profile}, we compared the modeled and observed
synchrotron brightnesses and photon indices predicted by each of the
the synchrotron losses or magnetic damping models. We found a good
match to the X-ray data in terms of shape profiles as well as in
absolute normalization.  This allowed us to constrain the injection
efficiency, $\eta_{\mathrm{inj}}$, the immediate post-shock magnetic
field, $B_2$, the ratio between the electron and proton distributions
at relativistic energies, $K_{\mathrm{ep}}$ and the maximum energy
reached by the electrons, $E_{\mathrm{e,max}}$.  On the other hand,
for neither model did we find very good agreement to the radio data in
terms of the absolute normalizations or shape of the rim profiles.
However, there are several model assumptions on which our results and
numerical values depend.  Here we detail these dependencies,
specifically how the limitations of the particle acceleration model
(used to calculate the synchrotron emission properties) may impact our
results, how important is the shape of the cutoff in the electron
spectrum in our modeling, and how a better model for the magnetic
damping can change the modeled profiles.

There are uncertainties associated with the simple acceleration model
\citep{bee99} we use to generate the particle distributions at the
shock.  In contrast to more exact Monte Carlo or kinetic shock models
calculations, the simple model we employ here provides an analytical
approximation to the particle spectrum that consists of broken
power-laws with three slopes characterizing the low, intermediate and
high energy regimes.  Furthermore, the energy at which the slope
changes at low energy is forced to be at $E = m_{\mathrm{p}} \: c^2
\sim 1 \; \mathrm{GeV}$, which corresponds generally to the energy
where the high energy electrons emit in the radio.  More accurate
models for the particle spectrum from diffusive shock acceleration
\citep[which produce smooth particle spectra, see][]{bl02}, tend to
predict a relatively higher number density of radio-emitting electrons
(by a factor of a few) around 1 GeV than at higher or lower energies
compared to the approximate particle acceleration model used here.
This will result in a higher radio synchrotron flux in both the
magnetic damping and synchrotron losses model cases. However, it is
difficult to quantify the amount of increase without employing these
different calculations for the particle spectra, which is beyond the
scope of this study.

\begin{figure*}[t]
\centering
\includegraphics[width=14cm]{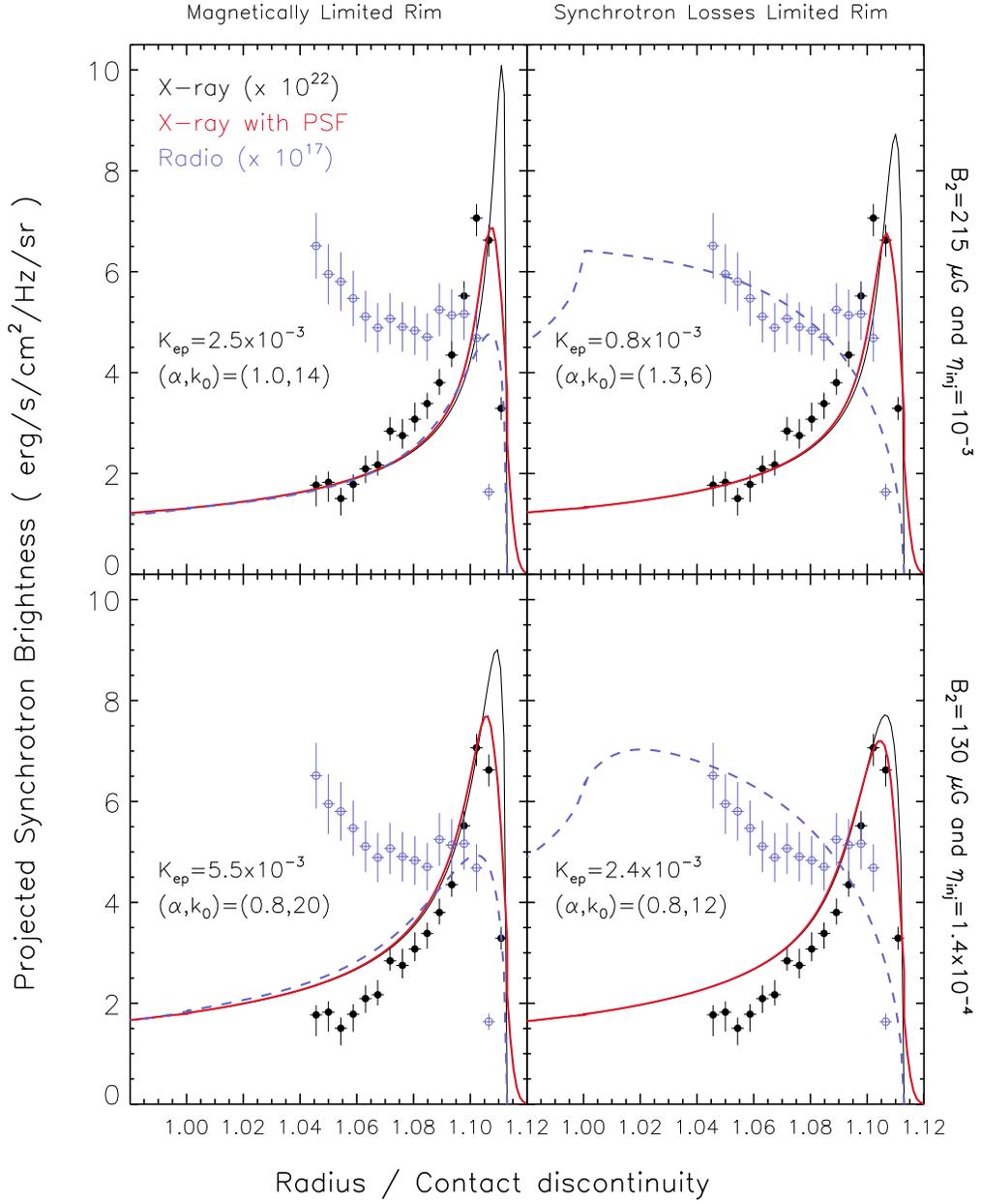}
\caption{Same as Figure \ref{fig_synchproj_profile} but
with a varying shape of the cutoff in the electron spectrum
(i.e., where $\alpha$ is a free parameter in Eq. \ref{fe_alpha}).}
\label{fig_synchproj_profile_beta}
\end{figure*}

There is also uncertainty associated with the cutoff of the particle
distribution function at high energy where the relativistic electrons
emit X-rays.  All previous results in this work were presented
assuming a purely exponential cutoff in the accelerated particle
spectrum (see \S \ref{subsect-CR-hydro+spectra}).  However, deviations
from homogeneity can cause the cutoff to be narrowed or broadened
\citep{pe06}.  And in fact integrated-spectral fits to the synchrotron
X-rays from SNRs often require that the cut-off be broadened
\citep{els01}. Thus we are led to investigate how dependent our
results are to the shape of the electron cutoff.  For that purpose, we
consider a new CR electron spectrum:
\begin{equation} \label{fe_alpha}
f_{\mathrm{e}}(E) = a \: K_{\mathrm{ep}} \: E^{-\Gamma(E)} \:
 \exp \left( - \frac{1}{\alpha} \left[
 \frac{E}{E_{\mathrm{e,max}}} \right]^{\alpha} \right),
\end{equation}
where $\alpha$ is a number characterizing the shape of the cutoff
\citep[see][]{elb00}.  It appears that any variations in $\alpha$ will
strongly impact the ratio of X-ray to radio fluxes as shown in Figure
\ref{fig_synchproj_profile_beta}. This is because most of the
synchrotron radiation is emitted right at the blast wave, i.e., where
losses by synchrotron cooling downstream (which erase any information
on the shape of the cutoff) do not have time to modify the spectrum of
the accelerated electrons.  For instance, in the magnetic damping case
when $\eta_{\mathrm{inj}} = 1.4 \times 10^{-4}$ and $B_2 = 130 \:
{\mu}\mathrm{G}$, values of $\alpha = 0.8$, $k_0=20$ and
$K_{\mathrm{ep}}=5.5 \times 10^{-3}$ provide the right radio and X-ray
fluxes at the rim (bottom-left panel) and an averaged X-ray photon
index of 2.9 that is consistent with the observations.  In the
synchrotron losses case, we found that this is obtained when $\alpha =
1.3$, $k_0 = 6$ and $K_{\mathrm{ep}}=0.8 \times 10^{-3}$ when
$\eta_{\mathrm{inj}} = 10^{-3}$ and $B_2 = 215 \: {\mu}\mathrm{G}$
(top-right panel), and $\alpha = 0.8$, $k_0 = 12$ and
$K_{\mathrm{ep}}=2.4 \times 10^{-3}$ when $\eta_{\mathrm{inj}} = 1.4
\times 10^{-4}$ and $B_2 = 130 \: {\mu}\mathrm{G}$ (bottom-right
panel).

Of course, these values will change depending on the quality of the
data.  While there is little systematic uncertainty associated with
the X-ray flux, the radio flux can be systematically off by a factor
of order 2 (see \S \ref{subsect-radio-xray-confidence}).  An
underestimate (overestimate) of the observed radio flux would increase
(decrease) the $K_{\mathrm{ep}}$ values by the same factor in the
models where $\alpha$ is a free parameter.  This, in turn, increases
(decreases) the modeled X-ray flux. To make it consistent with the
observed X-ray flux requires modifying the parameters characterizing
the cutoff in the electron spectrum (i.e., $E_{\mathrm{e,max}}$ or
$k_0$ and $\alpha$).  An increase (decrease) of the observed radio flux
by a factor of 2 for the same X-ray flux and average photon index
requires that we roughly lower (increase) $k_0$ by $\sim 50\%$ and
increase (lower) $\alpha$ by $\sim 20\%$ in the magnetic damping and
synchrotron losses model cases (assuming $\eta_{\mathrm{inj}} = 1.4
\times 10^{-4}$ and $B_2 = 130 \: {\mu}\mathrm{G}$).

The point of the preceding exercise is to show that the X-ray/radio
flux ratio is sensitive to the detailed cutoff of the particle energy
spectrum, about which we have few independent observational
constraints.  While we attach little importance to the precise values
of $\alpha$ (and other parameters) derived here it is comforting to
note that they fall within generally accepted ranges.  On the other
hand this modest little study demonstrates that the X-ray/radio flux
ratio by itself has little power to constrain the model parameters.

Finally, we consider the description of the magnetic damping model
proposed by \cite{poy05}.  This model assumes a phenomenological
exponential falloff in magnetic field strength with a characteristic
length, $l_{\mathrm{d}}$, given by various possible damping mechanisms
(see Appendix \ref{app-MF-damped}).  However, there is no reason why
the dropoff should not be faster or slower.  A slower (faster)
magnetic field decay behind the shock would result in a projected
profile of the synchrotron emission which is less (more) peaked in the
radio.  A slower decrease could be obtained if rather than the
amplified magnetic field, it is for instance the amplified magnetic
energy that is exponentially damped downstream of the shock on the
spatial scale $l_{\mathrm{d}}$. A further complication can be that the
parallel and perpendicular components of the magnetic field are damped
on different length scales. Taking into account those possibilities
and the fact that the magnetic field orientation may not be negligible
for the calculation of the synchrotron emissivity may change the
predictions of the magnetic damping model. (Note that the effects due
to magnetic field orientation may impact the prediction of the
synchrotron losses model case as well).

%*******************************************
\section{Conclusion\label{sect-conclusion}}
%*******************************************

The present paper addresses questions concerning the heating of the
ambient gas and acceleration of relativistic particles at SNR
blast waves. In young ejecta-dominated SNRs, the blast wave appears in
the form of an outer geometrically thin rim where most of the
synchrotron X-ray emission is confined. This provides strong evidence
for the production and acceleration of cosmic-ray electrons to very
high energies, right at the shock. Because in theory the blast wave
compresses and heats the ambient gas to very high temperatures, a
thermal X-ray component is also expected, but yet, there is little to
no evidence for such component. The region between the blast wave and
the shocked ejecta interface (or contact discontinuity) is actually
X-ray dark. The physics associated with collisionless shocks in SNRs
is not well understood. Where is the shocked ambient medium? What is
the fundamental physical mechanism for the production of the thin
X-ray synchrotron emitting rims? These are precisely the outstanding
questions that we address.

The best target for such study is probably the Tycho SNR as
observed by the \textit{Chandra X-ray Observatory}. The
quasi-circularity and regularity of the X-ray synchrotron emitting
rims in Tycho provide a very convenient framework for a combined
observational and theoretical investigation. Our starting point
was the X-ray analysis of the region between the blast wave and
contact discontinuity, which is well resolved in Tycho. In several
azimuthal regions, we extracted a set of spectra (between 0.7 and
7 keV) over several arcseconds as one moves in radially from the
blast wave to the contact discontinuity. These spectra contain
information on the thermal and nonthermal populations that can be
extracted provided that we can separate their respective
contributions to the X-ray emission. The radial variations of the
X-ray spectrum indicate a dominant contribution from and a
softening of the synchrotron component behind the blast wave. We
found in particular that the spectral index of the synchrotron
component increases from a value of $\sim 2.6$ at the blast wave
to $\sim 3.0$ behind the bright X-ray rim. The radial profiles of
the X-ray synchrotron emission were compared to similar profiles
in the radio band. These profiles rise at the blast wave in a very
similar manner but while the X-ray emission drops rapidly the
radio profile tends to remain (more or less) at its peak value.
The radial variations of the X-ray spectrum indicate also another
contribution (primarily Si and S line emission) from small knots
of shocked ejecta that have nearly reached the blast wave.

The lack of thermal contribution from the shocked ambient medium to
the X-ray spectrum implies, in the most general case, that the shocked
ambient gas has either a low pre-shock density ($\sim 0.2 \;
\mathrm{cm}^{-3}$) with no constraint on its temperature or somewhat
higher density but with a temperature below 1 keV. To go further, we
built an emission model for the shocked ambient medium based on
cosmic-ray hydrodynamic models which satisfy the observed ratio of
radii ($\sim 1.1$) between the blast wave and the contact
discontinuity and the expansion measurements. Spectral analysis using
this model indicates that the ambient medium density must be lower
than $0.3 \: \mathrm{cm}^{-3}$, assuming a kinetic energy of the
explosion of $10^{51} \: \mathrm{ergs}$.  Systematic errors due to,
for example, a more astrophysically appropriate initial ejecta density
profile could push this limit to $\sim 0.6\: \mathrm{cm}^{-3}$.
Higher densities lead to X-ray emission that cannot be hidden by
interstellar absorption. We found that even though the cosmic-ray
hydrodynamic models predict that the shocked ambient gas is much less
hot than in a pure gas shock, it is never sufficiently cool that its
emission gets shifted to the extreme UV range. This does not seem to
be a viable explanation for the lack of thermal X-ray emission from
the shocked ambient medium.

Much of our effort in this paper went toward modeling the intensity
profiles in the radio and X-ray bands and the X-ray spectral
variations of the synchrotron emission at the rim. Our goal was to
determine whether the observed X-ray rims reflect the spatial
distribution of the highest energy electrons or that of the magnetic
field. The most critical ingredient in the modeling is the magnetic
field and we have considered two scenarios for its post-shock
evolution, assuming that it has been already amplified at the blast
wave: one where the magnetic field is simply advected downstream from
the shock and remains relatively high in the post-shock region, and
one where the magnetic field is rapidly decreasing behind the shock
because of the damping or relaxation of the turbulence. We refer to
these as the synchrotron losses case and magnetic damping case,
respectively. In both cases, a model with a cosmic-ray injection of $3
\times 10^{-4}$, a post-shock amplification of the magnetic field up
to $175 \: {\mu}\mathrm{G}$ and an electron-to proton density at
relativistic energies of order $10^{-3}$ accurately reproduces the
narrow gap between the blast wave and contact discontinuity, the width
and brightness of the X-ray synchrotron rims. This model assumes an
ambient density of $0.2 \: \mathrm{cm}^{-3}$ and a kinetic energy of
the explosion of $10^{51} \: \mathrm{ergs}$. In addition, both
synchrotron losses and magnetic damping scenarios produce radial
photon index variations that can accommodate the range of observed
variations seen in the \textit{Chandra} data. This is because, even in
the magnetic damping case, synchrotron losses play a role in shaping
the X-ray morphology and spectral index variations at the rim. A good
match to the average X-ray photon index is obtained when relaxing the
Bohm diffusion assumption. This constrains the diffusion coefficient
to be $\sim 7-10$ times the Bohm value and implies a maximum energy of
the electrons of $10 \; \mathrm{TeV}$.  Right at the shock, this would
correspond to a cutoff energy in the synchrotron spectrum of $0.3 \:
\mathrm{keV}$.

The grossest difference between the magnetic damping and synchrotron
losses model cases concerns the radio synchrotron emission.  Given
possible systematic uncertainties associated with the absolute radio
flux, the theoretical description of the particle distribution
functions, the model for the magnetic field decay behind the shock,
the possibility of a broadening in the high-energy cutoff of the
accelerated particle spectrum, and potentially the projection of the
synchrotron emissivity onto the line-of-sight, we cannot reject one
model versus the other based on the ratio of absolute X-ray and radio
intensities.  Nevertheless, we can use the shapes of the projected
synchrotron emission.  As regards the profile of the radio emission,
the magnetic damping case produces a sharp rise in brightness at the
blast wave, as observed, which the synchrotron cooling profile fails
to do. One possible source of uncertainty in this comparison comes
about because our X-ray and radio observations were made at widely
separated times and the relative positioning of the rims in the two
wave bands is subject to error since the remnant's angular expansion
rate is still only poorly known.  On the other hand, the synchrotron
cooling profile yields a gradually rising radio profile behind the
blast wave, a feature that is not in contradiction with the
observation.  Perhaps the actual situation is a combination of these
two scenarios: a geometrically thin region of enhanced magnetic field
right at the shock, that is only partially damped to some intermediate
field value (i.e., $50-100 \: {\mu}\mathrm{G}$ rather than the $5 \:
{\mu}\mathrm{G}$ value we assumed here) in the post-shock zone or
where additional magnetic field has been generated by turbulent
motions caused by the outermost pieces of ejecta (which we see in the
X-ray data).  This type of magnetic field configuration might also be
more consistent with the radio profiles in the NE and NW rim regions
(see Fig.~11).  Coming up with realistic evolutionary scenarios in
these advanced cases will require some care, but they may provide
interesting constraints on the mysterious processes by which magnetic
fields are generated at collisionless shock waves.

As we have shown in this article, a detailed study of the X-ray and
radio emissions behind the SNR blast wave is crucial for understanding
the nature of high Mach collisionless shocks.  And much more remains
to be done. Given the importance of the relative radio and X-ray rim
morphologies, a key issue involves having radio and X-ray observations
taken at roughly the same time, so that uncertainties due to the
remnant's expansion can be minimized.  Of equal importance is
obtaining a reliable flux-calibrated map of the radio emission so that
the point-to-point relationship between radio and X-ray synchrotron
emission can be established and used to discriminate between
models. The X-ray rims of Tycho, as observed by \textit{Chandra}, are
still largely unresolved.  Reobservation with some portion of the rim
at the prime, on-axis pointing location of \textit{Chandra} with the
narrowest PSF (rather than some $4\arcmin$ off-axis where the rims are
in the current data set and the PSF is some $2\arcsec$) would help
better determine their structure. Other young SNRs, observed by
\textit{Chandra}, are without question suitable for similar studies
along the lines of what we have done here.  Finally a deeper
understanding of the magnetic field and synchrotron emission
properties of the forward shock in Tycho would benefit from
theoretical investigations using more sophisticated numerical
hydrodynamical models of cosmic-ray modified shocks
\citep[e.g.,][]{elc05}.

\acknowledgments

We acknowledge D.~C. Ellison for many discussions on particle
acceleration. We thank the referee for his careful reading of the
paper and judicious comments. GCC would like to thank M. Pohl, H. Yan
and A. Lazarian for their inspiring paper on magnetic filaments.  GCC
acknowledges J.~S. Warren for providing the inputs needed from her PCA
of the Tycho \textit{Chandra} data, C. Badenes for having proposed the
use of his model and U. Hwang and R. Petre for the discussions
preceding this project. JPH thanks J. Dickel and E.  Reynoso for
providing the VLA image and for helpful discussions about the radio
emission from Tycho. Financial support was provided by NASA grant
NNG05GP87G and \textit{Chandra} grants GO4-5076X and GO6-7016B to
Rutgers, The State University of New Jersey.

%===========================================================

\appendix

%*******************************************
\section{Evolution and profile of the magnetic field\label{app-MF-profiles}}
%*******************************************

\subsection{Damped magnetic field\label{app-MF-damped}}

To obtain a magnetic filament as suggested by \cite{poy05}, we use
the following phenomenological magnetic field profile
to describe the evolution of the magnetic field in a given fluid element:
\begin{equation}\label{Bexpo-evol}
 B(r) = B_{\infty} + ( B_2 - B_{\infty} ) \: \exp \left( - \frac{r_{\mathrm{s}} - r}{l_{\mathrm{d}}} \right),
\end{equation}
where $r$ is the position of a fluid element and $r_{\mathrm{s}}$ the
shock radius at the same time. $B_2$ is the
immediate post-shock magnetic field which later decays to
$B_{\infty}$ with a characteristic length $l_{\mathrm{d}}$. We set
$B_{\infty}$ to the value of $5 \: \mu{\mathrm{G}}$ as done in \cite{poy05}.
The length
$l_{\mathrm{d}}$ is equal to $\max\{ l_{\mathrm{k}},
l_{\mathrm{A}}, l_{\mathrm{f}}\}$ where $l_{\mathrm{k}},
l_{\mathrm{A}}$ and $l_{\mathrm{f}}$ are different damping lengths
given by:
\begin{eqnarray}\label{lk}
l_{\mathrm{k}} & = & \frac{5}{\pi} \: \frac{u_j}{c_{\mathrm{A}}} \: \lambda,
\\ \label{lA} l_{\mathrm{A}} & = & \frac{1}{2 \: \sqrt{2 \: \pi}} \:
\frac{u_j}{c_{\mathrm{A}}} \: \sqrt{ \lambda \: L }, \\ \label{lf}
l_{\mathrm{f}} & = & \frac{1}{2 \: \sqrt{2 \: \pi}} \: \frac{u_j \:
v_{\phi}}{v_{\mathrm{L}}^2} \: \sqrt{\lambda \: L},
\end{eqnarray}
where $u_j$ is the downstream flow speed (in the shock frame),
$c_{\mathrm{A}} = B_2 / \sqrt{ 4 \: \pi \: \rho_2}$ 
the Alfv\'en speeds just behind the blast wave (with $B_2$ and $\rho_2$ the
magnetic field and mass density just behind the shock, respectively),
$v_{\mathrm{L}}$ the turbulence velocity at the injection scale
(typically a few hundreds of km/s)
and $v_{\phi}$ a velocity equal to the Alfv\'en speed
for the high-$\beta$ plasma considered here ($\beta$ being
the pressure ratio between the gas and the magnetic field),
$\lambda$ the wavelength of the
turbulent magnetic field of order the Larmor radius, $r_{\mathrm{L}} = E_{\mathrm{max}}/(e \: B_2)$, of the maximum energy protons and 
$L$ the outer turbulence scale of order the shock radius $r_{\mathrm{s}}$. 
Since $\lambda \ll L$, we have $l_{\mathrm{k}} < l_{\mathrm{A}}$
and, assuming that $v_{\mathrm{L}} \sim v_{\phi}$, we have 
$l_{\mathrm{f}} \sim l_{\mathrm{A}}$. This yields 
$l_{\mathrm{d}} = l_{\mathrm{A}} \propto B_2^{-3/2}$ so
the higher the strength of the magnetic field just behind the shock,
the smaller the damping length.
Note that the above damping lengths have been slightly modified or
rearranged compared to the formula given by \cite{poy05}.

\subsection{Advected magnetic field\label{app-MF-advected}}

We assume that the magnetic field is simply carried by the flow,
frozen in the plasma, so that the parallel and perpendicular magnetic
field components, separately, evolve conserving flux.  Evolution
equations are fully described in \cite{cad05} and references therein.

%*******************************************
\section{Slope of particle spectrum after energy losses\label{app-slope-losses}}
%*******************************************

Let us assume that the particle differential spectrum
$N_\mathrm{s}$ can be locally described with a power-law with an
index $\Gamma_{\mathrm{s}}$ around an energy $E_\mathrm{s}$, at a
time $t_\mathrm{s}$: $ N_\mathrm{s} \equiv \mathcal{N}(
E_\mathrm{s} ) = K_\mathrm{s} \:
E_\mathrm{s}^{-\Gamma_{\mathrm{s}}}$.

Due to the adiabatic expansion and radiative losses, the particle
spectrum will be modified in terms of energy ($E_\mathrm{s}
\rightarrow E$) and density ($N_\mathrm{s} \rightarrow N$). We
assume that this modified particle spectrum can be still described
with a power-law with an index $\Gamma$, at a later time $t$: $ N
\equiv \mathcal{N}( E ) = K \: E^{-\Gamma}$.

Independently of the shape of the particle spectrum, the change in
energy and density is given by \citep{re98}:
\begin{eqnarray}\label{E-E0}
 E &=& \alpha^{1/3} \: \frac{E_\mathrm{s}}{1 + \Theta \: E_\mathrm{s}} \\ \label{N-N0}
 N &=& N_\mathrm{s} \: \frac{1}{\alpha} \: \frac{dE_\mathrm{s}}{dE} = N_\mathrm{s} \: \alpha^{-2/3} \: \left( \frac{E_\mathrm{s}}{E} \right)^2
\end{eqnarray}
where $1 / \alpha \equiv V / V_\mathrm{s}$ is the relative change
in volume between time $t$ and time $t_\mathrm{s}$, and $\Theta$
a radiative loss term that includes both synchrotron and inverse Compton on the
radiation field:
\begin{equation}\label{theta}
 \Theta \equiv \int_{t_\mathrm{s}}^{t} \: a \:
 B_{\mathrm{eff}}^2(\tau) \: \alpha^{1/3}(\tau) \: d\tau
\end{equation}
with $a = 4 \: e^4 / ( 9 \: m^4 c^7 )$ a constant depending on the
particle mass $m$, and $B_{\mathrm{eff}} \equiv \left( B^2 +
B_{\mathrm{cbr}}^2\right)^{1/2}$ an effective magnetic field which
includes the magnetic field inside the remnant, $B$, and the
magnetic field with energy density equal to that in the radiation
field, $B_{\mathrm{cbr}}$. If this is the microwave background
then $B_{\mathrm{cbr}} = 3.27 \: \mu{\mathrm{G}}$.

The change in slope of the particle spectrum is obtained by using
Eqs (\ref{E-E0}) and (\ref{N-N0}):
\begin{equation}\label{slopeG}
 \Gamma \equiv -\frac{d \: \ln N}{d \: \ln E} = \left( \Gamma_\mathrm{s}
 - 2 \right) \: \left[ 1 + \Theta \: E_\mathrm{s} \right] + 2.
\end{equation}

From Eq. (\ref{slopeG}), we see that adiabatic expansion only
(i.e., $\Theta = 0$) does not cause any change in the slope of the
particle spectrum, contrary to radiative losses.

%*******************************************
\section{Projected photon index profile\label{app-ph-index-profile}}
%*******************************************

The projection along the line-of-sight of a radial emissivity
profile $\mathcal{E}_{\nu}$ results in a brightness profile
$\mathcal{B}_{\nu}$ of the form:
\begin{equation}\label{B}
 \mathcal{B}_{\nu}(\rho) = 2 \: R_{s} \: \int_{0}^{\ell} \:
 \mathcal{E}_{\nu}(r) \: dz \; \; \mathrm{with} \; \;
 \left\{ \begin{array}{ccc}
  r^2 & = & \rho^2 + z^2 \\
 \ell^2 & = & 1 - \rho^2 \\
 \end{array} \right. ,
\end{equation}
where $r$ is the distance to the center of a sphere, $\rho$ the
distance between the center of the disk (i.e., projection of the
sphere onto a plane) and the line-of-sight, and $\ell$ the length
of the line-of-sight. All quantities are expressed in units of the
sphere's radius $R_{s}$.

If the emissivity decreases from its maximum
$\mathcal{E}_{\nu,\mathrm{max}}$ with a characteristic width
$a_{\nu}$ (in units of $R_{s}$), \cite{ba06} has demonstrated that
the brightness profile near the edge of the disk has the general
form:
\begin{equation}\label{Bedge}
\mathcal{B}_{\nu}(\rho) \simeq 2 \: R_{s} \: \sqrt{2 \: a_{\nu}}
\: \mathcal{E}_{\nu,\mathrm{max}} \; g(y_{\nu}),
\end{equation}
where $g$ is a functional form and $y_{\nu}
\equiv(1-\rho)/a_{\nu}$. In the case of an exponentially
decreasing emissivity profile:
\begin{equation}\label{g}
 g(x) = e^{-x} \: \int_{0}^{\sqrt{x}} \: e^{u^2} \: du.
\end{equation}
The maximum of $g$ occurs at $y_{\nu}^{0} \simeq 0.854$.

In the same limit of small $a_{\nu}$, the brightness toward the
center is:
\begin{equation}\label{B0}
 \mathcal{B}_{\nu}(0) \simeq 2 \: R_{s} \: a_{\nu} \:
 \mathcal{E}_{\nu,\mathrm{max}}.
 \:
\end{equation}

From Eq. (\ref{Bedge}), one can derive the slope $\alpha_{\nu}$ (at
given frequency $\nu$) of the projected spectrum built from the
projected brightness profile:
\begin{equation}\label{alpha-nu}
\alpha_{\nu}(\rho) \equiv - \frac{d \ln \mathcal{B}_{\nu}}{d \ln
\nu} = - \frac{1}{2} \frac{d \ln a_{\nu}}{d \ln \nu} - \frac{d \ln
\mathcal{E}_{\nu,\mathrm{max}}}{d \ln \nu} - \frac{d \ln
g(y_{\nu})}{d \ln \nu}
\end{equation}

In the exponential case,
\begin{equation}\label{dlng}
\frac{d \ln g(y_{\nu})}{d \ln \nu} = \frac{1}{2} \frac{d \ln
a_{\nu}}{d \ln \nu} \: h(y_{\nu}) \; \; \mathrm{with} \; \; h(x) =
2 \: x - \frac{\sqrt{x} }{g(x)}.
\end{equation}
Then, the slope profile $\alpha_{\nu}$ reaches its maximum at the
radius $\rho^{\star} = 1 - a \: y_{\nu}^{\star}$ \ obtained by
solving:
\begin{equation}\label{dh}
 \frac{d\alpha_{\nu}}{d \rho} = 0 \Leftrightarrow \frac{dh(y_{\nu}^{\star})}{d \rho} =
 0 \Leftrightarrow \frac{2}{a_{\nu}} \: k(y_{\nu}^{\star}) = 0 \; \; \mathrm{where} \; \;
k(x) = -1 + \frac{1 + h(x)}{4 \: \sqrt{x} \: g(x)}.
\end{equation}
We find $y_{\nu}^{\star} \simeq 4.386$. For comparison, the
brightness profile decreases inwards to half its maximum value at
$y_{\nu}^{1} \simeq 4.685$.

The maximum value of the photon index and
its value at the edge (using Eqs \ref{alpha-nu}-\ref{dlng} and $\displaystyle{\lim_{x\rightarrow 0}} h(x) = -1$)
and at the center (see Eq. \ref{B0}) are respectively:
\begin{eqnarray}\label{Gph-nu}
\Gamma_{\nu}(\rho^{\star}) & = 1 + \alpha_{\nu}(\rho^{\star}) & = 1 -
\left( \frac{1 + h(y_{\nu}^{\star})}{2} \right) \: \frac{d \ln
a_{\nu}}{d \ln \nu} - \frac{d \ln
\mathcal{E}_{\nu,\mathrm{max}}}{d \ln \nu}, \\
\Gamma_{\nu}(1) & = 1 + \alpha_{\nu}(1) & = 1 - \frac{d \ln
\mathcal{E}_{\nu,\mathrm{max}}}{d \ln \nu}, \\
\Gamma_{\nu}(0) & = 1 + \alpha_{\nu}(0) & = 1 - \frac{d \ln a_{\nu}}{d \ln \nu} - \frac{d \ln
\mathcal{E}_{\nu,\mathrm{max}}}{d \ln \nu},
\end{eqnarray}
with $h(y_{\nu}^{\star}) \simeq 1.370$.

If $a_{\nu} \propto 1 / \sqrt{\nu}$ (as expected if the rims are
limited by synchrotron losses), we find
$\Gamma_{\nu}(\rho^{\star}) - \Gamma_{\nu}(1) \simeq 0.593$ and
$\Gamma_{\nu}(\rho^{\star}) - \Gamma_{\nu}(0) \simeq 0.093$.

%===========================================================

%++++++++++++++++++++++++++++++++++++++++++++++++++++
% Bibliographie a cet endroit ( style A&A )
\bibliographystyle{aa} % A&A style
\bibliography{ms} % ne surtout pas mettre le .bib !!!
%+++++++++++++++++++++++++++++++++++++++++++++++++++++

%#######################################################
% FIGURES
%#######################################################

\clearpage

\clearpage

\clearpage

\clearpage

\clearpage

\clearpage

\clearpage

\clearpage

\clearpage

\clearpage

\clearpage

\clearpage

\clearpage

\clearpage

\clearpage

\clearpage

\clearpage

\clearpage

%#######################################################
% TABLES
%#######################################################

\begin{table*}[t]
\caption{Geometrical parameters of the selected sector regions
(see Fig. \ref{fig_Tycho}). Angles are counted counterclockwise
and zero starts from west. The center of each region is chosen in
order to match the local curvature of the rims. For comparison,
the nominal center of expansion in the X-rays is
$\alpha_{\mathrm{J2000}}$ = $00$h$25$m$19$s,
$\delta_{\mathrm{J2000}}$ = $64^\circ 08\arcsec 10\arcsec$
\citep[][ROSAT data]{hu00} and the center that minimizes the
ellipticity of the blast wave is $\alpha_{\mathrm{J2000}}$ =
$00$h$25$m$19.4$s, $\delta_{\mathrm{J2000}}$ = $64^\circ 08\arcsec
14.0\arcsec$ \citep{wah05}. The last column gives the reference
radial bin position of the blast wave (BW) that we chose for each
rim.} \label{tab-regions} \centering
\begin{tabular}{llllcccccc}
\tableline \tableline
\multicolumn{2}{l}{Region} & \multicolumn{2}{c}{Region Center of Curvature} & \multicolumn{2}{c}{Angle} & \multicolumn{2}{c}{Radius} & \multicolumn{1}{c}{Position}\\
 & & $\alpha_{\mathrm{J2000}}$ & $\delta_{\mathrm{J2000}}$ & ${\min}$ & ${\max}$ & ${\min}$ & ${\max}$ & BW \\ \tableline
 & \multicolumn{1}{l}{Ejecta} & & & & & $220\arcsec$ & $230\arcsec$ & \\
W & \multicolumn{1}{l}{Rim} & $00$h$25$m$19.5$s & $64^\circ 08\arcmin 14.0\arcsec$ & $335^\circ$ & $360^\circ$ & $240\arcsec$ & $264\arcsec$ & $255.5\pm0.5\arcsec$ \\
 & \multicolumn{1}{l}{Background} & & & & & $276\arcsec$ & $294\arcsec$ & \\ \tableline
 & \multicolumn{1}{l}{Ejecta} & & & & & $178\arcsec$ & $188\arcsec$ & \\
NW & \multicolumn{1}{l}{Rim} & $00$h$25$m$13.0$s & $64^\circ 08\arcmin 11.0\arcsec$ & $46^\circ$ & $59^\circ$ & $198\arcsec$ & $222\arcsec$ & $214.5\pm0.5\arcsec$ \\
 & \multicolumn{1}{l}{Background} & & & & & $234\arcsec$ & $252\arcsec$ & \\ \tableline
 & \multicolumn{1}{l}{Ejecta} & & & & & $562\arcsec$ & $572\arcsec$ & \\
NE & \multicolumn{1}{l}{Rim} & $00$h$24$m$52.5$s & $64^\circ 03\arcmin 10.0\arcsec$ & $111^\circ$ & $119^\circ$ & $582\arcsec$ & $606\arcsec$ & $599.5\pm0.5\arcsec$ \\
 & \multicolumn{1}{l}{Background} & & & & & $618\arcsec$ & $636\arcsec$ & \\ \tableline
\end{tabular}
\end{table*}

\clearpage

\begin{table}[t]
\caption{Best-fit blast wave (BW) radii and shell widths as
obtained from different emissivity projected shell models
convolved with a \textit{Chandra} PSF. The blast wave radii are
consistent with the radial bin that we chose as the blast wave
position in Table \ref{tab-regions}.} \label{tab-radii-psf}
\centering
\begin{tabular}{lccccl}
\tableline \tableline
\multicolumn{1}{c}{Rim} & \multicolumn{2}{c}{Uniform emissivity} & \multicolumn{2}{c}{Exponential emissivity} & \multicolumn{1}{c}{Fit quality} \\
 & \multicolumn{1}{c}{BW Radius} & \multicolumn{1}{c}{Width} & \multicolumn{1}{c}{BW Radius} & \multicolumn{1}{c}{Width} & \\ \tableline
W & $255.1 \pm 0.2 \arcsec$ & $< 0.6 \arcsec$ & $255.0 \pm 0.2 \arcsec$ & $< 0.6 \arcsec$ & Good \\
NW & $ 214.5 \pm 0.2 \arcsec$ & $< 0.4 \arcsec$ & $214.4 \pm 0.2 \arcsec$ & $< 0.1 \arcsec$ & Bad \\
NE & $599.2 \pm 0.2 \arcsec$ & $< 0.4 \arcsec$ & $599.3 \pm 0.2
\arcsec$ & $< 0.2 \arcsec$ & Bad \\ \tableline
\end{tabular}
\end{table}

\clearpage

\begin{table*}[t]
\caption{Best-fit photon index $\Gamma_{\mathrm{ph}}$ and $\chi^2$
for different spectral models as a function of position behind the
blast wave in the rim W: 1 - power-law with a varying absorption,
2 - power-law with a fixed absorption, 3 - power-law plus a
template for the shocked ejecta with a fixed absorption. The
hydrogen column density is $\mathrm{N}_{\mathrm{H}} =
\mathrm{N}_{\mathrm{H},22} \times 10^{22}$ cm$^{-2}$. The errors
are in the range $\Delta \chi^2 < 2.7$ (90\% confidence level) on
one parameter and are given only when $\chi^2 / \mathrm{dof} <
2$.} \label{tab-slope-xspec-fit} \centering
\begin{tabular}{cllllll}
\tableline \tableline
\multicolumn{1}{r}{Rim W} & \multicolumn{2}{c}{1 - power-law} & \multicolumn{2}{c}{2 - power-law} & \multicolumn{2}{c}{3 - power-law + ejecta template} \\
& \multicolumn{2}{c}{(free $\mathrm{N}_{\mathrm{H}}$)} & \multicolumn{2}{c}{($\mathrm{N}_{\mathrm{H},22} = 0.7$)} & \multicolumn{2}{c}{($\mathrm{N}_{\mathrm{H},22} = 0.7$)} \\
Position & \multicolumn{1}{c}{$\Gamma_{\mathrm{ph}}$} &
\multicolumn{1}{c}{$\chi^2$ (dof)} &
\multicolumn{1}{c}{$\Gamma_{\mathrm{ph}}$} &
\multicolumn{1}{c}{$\chi^2$ (dof)} &
\multicolumn{1}{c}{$\Gamma_{\mathrm{ph}}$} &
\multicolumn{1}{c}{$\chi^2_{\mathrm{REF}}$ (dof)} \\ \tableline
$+1\arcsec$ & 2.89 & 80 (37) & 2.57 & 86 (38) & 2.52 & 79 (37) \\
$0$ & 2.67 (2.58-2.83) & 87 (94) & 2.59 (2.66-2.53) & 90 (95) & 2.57 (2.51-2.64) & 91 (94) \\
$-1\arcsec$ & 2.61 (2.52-2.69) & 180 (145) & 2.57 (2.62-2.52) & 180 (146) & 2.55 (2.49-2.60) & 166 (145) \\
$-2\arcsec$ & 2.70 (2.62-2.79) & 160 (143) & 2.66 (2.71-2.62) & 161 (144) & 2.65 (2.60-2.70) & 150 (143) \\
$-3\arcsec$ & 2.88 (2.78-2.99) & 164 (121) & 2.78 (2.83-2.72) & 169 (122) & 2.77 (2.70-2.83) & 154 (121) \\
$-4\arcsec$ & 2.64 (2.53-2.76) & 141 (109) & 2.72 (2.79-2.66) & 143 (110) & 2.71 (2.64-2.78) & 131 (109) \\
$-5\arcsec$ & 2.84 (2.73-2.97) & 108 (98) & 2.83 (2.90-2.76) & 108 (99) & 2.82 (2.75-2.90) & 97 (98) \\
$-6\arcsec$ & 2.86 (2.72-3.01) & 109 (83) & 2.86 (2.94-2.78) & 109 (84) & 2.86 (2.75-2.96) & 101 (83) \\
$-7\arcsec$ & 2.79 (2.65-2.93) & 82 (76) & 2.89 (2.97-2.80) & 84 (77) & 2.89 (2.79-2.98) & 75 (76) \\
$-8\arcsec$ & 2.97 (2.82-3.13) & 92 (71) & 2.95 (3.04-2.87) & 92 (72) & 2.97 (2.84-3.08) & 69 (71) \\
$-9\arcsec$ & 2.86 (2.72-3.01) & 116 (72) & 2.94 (3.03-2.86) & 117 (73) & 2.96 (2.86-3.06) & 91 (72) \\
$-10\arcsec$ & 2.77 & 178 (75) & 2.81 & 178 (76) & 2.77 (2.65-2.89) & 85 (75) \\
$-11\arcsec$ & 2.88 & 306 (85) & 2.92 & 306 (86) & 2.90 (2.76-3.04) & 99 (85) \\
$-12\arcsec$ & 2.79 & 337 (84) & 2.79 & 332 (84) & 2.66 (2.53-2.81) & 96 (84) \\
$-13\arcsec$ & 2.95 & 415 (84) & 2.89 & 416 (85) & 2.82 (2.66-3.01) & 103 (84) \\
$-14\arcsec$ & 2.88 & 411 (84) & 2.94 & 413 (85) & 2.93 (2.77-3.12) & 126 (84) \\
$-15\arcsec$ & 2.93 & 417 (86) & 2.93 & 417 (87) & 2.93
(2.77-3.14) & 110 (86) \\ \tableline
\end{tabular}
\end{table*}

\clearpage

\begin{table*}[t]
\caption{Best-fit photon index $\Gamma_{\mathrm{ph}}$ and $\chi^2$
obtained with a model combining a power-law and a self-consistent
non-equilibrium (NEI) model built from a CR-hydro model with solar
abundances and $n_0 = 0.2 \: \mathrm{cm}^{-3}$, the absorption
being held fixed ($\mathrm{N}_{\mathrm{H}} = 0.7 \times 10^{22} \:
\mathrm{cm}^{-2}$). The NEI model aims to represent the shocked
ambient medium from the blast wave (position 0) to the contact
discontinuity (position $-25\arcsec$) in the rim W. $\Delta t$ is
the flow time and the volume $V_{52}$ is given in units of
$10^{52} \: D_{\mathrm{kpc}}^3 \: \mathrm{cm}^{3}$ where
$D_{\mathrm{kpc}}$ is the distance to the SNR in kpc. We show two
cases with different electron-to-proton temperature ratio
$\beta_{\mathrm{s}}$: 1 and 0.01.}
\label{tab-nei-hydro+ej-xspec-fit} \centering
\begin{tabular}{cllllll}
\tableline \tableline
\multicolumn{1}{r}{Rim W} & & & \multicolumn{4}{c}{power-law + ejecta template + CR-hydro NEI model} \\
 & \multicolumn{1}{c}{$V_{52}$} & \multicolumn{1}{c}{$\Delta t$} & \multicolumn{2}{c}{$\beta_{\mathrm{s}} = 1$} & \multicolumn{2}{c}{$\beta_{\mathrm{s}} = 0.01$} \\
Position & & (yr) & \multicolumn{1}{c}{$\Gamma_{\mathrm{ph}}$} &
\multicolumn{1}{c}{$\chi^2$ (dof)} &
\multicolumn{1}{c}{$\Gamma_{\mathrm{ph}}$} &
\multicolumn{1}{c}{$\chi^2$ (dof)} \\ \tableline
$0$   & $0.6$ & 10 & 2.59 (2.53- 2.67) & 84 (94) & 2.58 (2.50- 2.64) & 84 (94) \\
$-1\arcsec$ & $1.7$ & 28 & 2.57 (2.52- 2.62) & 160 (145) & 2.55 (2.50- 2.60) & 159 (145) \\
$-2\arcsec$ & $2.5$ & 46 & 2.68 (2.63- 2.73) & 149 (143) & 2.65 (2.60- 2.70) & 148 (143) \\
$-3\arcsec$ & $2.9$ & 64 & 2.81 (2.75- 2.87) & 159 (121) & 2.77 (2.71- 2.82) & 156 (121) \\
$-4\arcsec$ & $3.4$ & 81 & 2.77 (2.69- 2.84) & 123 (109) & 2.70 (2.62- 2.77) & 125 (109) \\
$-5\arcsec$ & $3.7$ & 97 & 2.89 (2.80- 2.97) & 89 (98) & 2.81 (2.71- 2.88) & 89 (98) \\
$-6\arcsec$ & $4.0$ & 113 & 2.94 (2.84- 3.03) & 93 (83) & 2.83 (2.72- 2.91) & 97 (83) \\
$-7\arcsec$ & $4.4$ & 130 & 2.99 (2.88- 3.10) & 69 (76) & 2.86 (2.75- 2.95) & 70 (76) \\
$-8\arcsec$ & $4.6$ & 147 & 3.09 (2.98- 3.22) & 52 (71) & 2.93 (2.83- 3.04) & 56 (71) \\
$-9\arcsec$ & $4.9$ & 164 & 3.08 (2.96- 3.20) & 82 (72) & 2.91 (2.80- 3.01) & 81 (72) \\
$-10\arcsec$ & $5.1$ & 180 & 2.87 (2.76- 3.01) & 83 (75) & 2.71 (2.61- 2.82) & 81 (75) \\
$-11\arcsec$ & $5.3$ & 196 & 3.02 (2.88- 3.19) & 108 (85) & 2.82 (2.70- 2.95) & 107 (85) \\
$-12\arcsec$ & $5.5$ & 211 & 2.75 (2.61- 2.92) & 97 (84) & 2.58 (2.46- 2.71) & 95 (84) \\
$-13\arcsec$ & $5.8$ & 226 & 2.95 (2.77- 3.17) & 115 (84) & 2.71 (2.56- 2.86) & 110 (84) \\
$-14\arcsec$ & $6.0$ & 241 & 3.08 (2.90- 3.30) & 131 (84) & 2.82 (2.67- 2.98) & 128 (84) \\
$-15\arcsec$ & $6.2$ & 256 & 3.05 (2.85- 3.27) & 116 (86) & 2.81 (2.67- 2.98) & 114 (86) \\
 \tableline
\end{tabular}
\end{table*}

\clearpage

\begin{table*}[t]
\caption{Immediate post-shock magnetic field $B_2$, blast wave
radius $R_{s}$, speed $V_{s}$ and distance to the remnant $D$
obtained from a CR-hydro model for different values of the
pre-shock ambient medium density $n_0$ (assuming an injection
efficiency $\eta_{\mathrm{inj}}$ of $10^{-3}$). We varied the
upstream magnetic field $B_0$ so that the blast wave to contact
discontinuity radii ratio was consistent with the one of the
observation in rim W, i.e., $R_{s}/R_{c}=1.113$. The high value of
the upstream magnetic field $B_0$ implicitly assumes that it has
been already significantly amplified by the CR-streaming
instability. In all cases, the overall and magnetic field
compression ratios are about $6$ and $5$, respectively. We give
the distance to the SNR, $D$, assuming an angular radius of $256
\arcsec$ for the blast wave. } \label{tab-CR+hydro-simu}
\centering
\begin{tabular}{ccccccrrr}
\tableline \tableline \multicolumn{1}{c}{$n_0$
($\mathrm{cm}^{-3}$)} & \multicolumn{1}{c}{$B_0$
($\mu{\mathrm{G}}$)} & \multicolumn{1}{c}{$B_2$
($\mu{\mathrm{G}}$)} & \multicolumn{1}{c}{$R_{s}$ (pc)} &
\multicolumn{1}{c}{$V_{s}$ (km/s)} & \multicolumn{1}{c}{$D$ (kpc)}
\\ \tableline
2.00 & 98 & 490 & 2.63 & 3420 & 2.12 \\
1.50 & 89 & 444 & 2.74 & 3570 & 2.21 \\
1.00 & 77 & 384 & 2.91 & 3780 & 2.34 \\
0.90 & 74 & 370 & 2.95 & 3840 & 2.38 \\
0.80 & 71 & 354 & 3.00 & 3900 & 2.42 \\
0.70 & 68 & 339 & 3.06 & 3980 & 2.47 \\
0.60 & 64 & 320 & 3.13 & 4070 & 2.52 \\
0.50 & 60 & 300 & 3.21 & 4170 & 2.59 \\
0.40 & 56 & 279 & 3.32 & 4310 & 2.67 \\
0.30 & 50 & 250 & 3.45 & 4490 & 2.78 \\
0.20 & 43 & 215 & 3.66 & 4760 & 2.95 \\
0.10 & 34 & 169 & 4.04 & 5250 & 3.26 \\
0.09 & 32 & 162 & 4.10 & 5330 & 3.31 \\
0.08 & 31 & 155 & 4.17 & 5420 & 3.36 \\
0.07 & 30 & 149 & 4.26 & 5530 & 3.43 \\
0.06 & 28 & 140 & 4.35 & 5650 & 3.50 \\
0.05 & 26 & 131 & 4.46 & 5800 & 3.60 \\ \tableline
\end{tabular}
\end{table*}

\end{document}